\documentclass[journal]{IEEEtran}
\usepackage{amsmath}
\usepackage{amssymb}
\usepackage{mathtools,amsthm}
\usepackage{comment}
\usepackage{graphicx}
\usepackage{caption}
\usepackage{subcaption}
\usepackage{hyperref}
\usepackage{xcolor}

\newtheorem{lemma}{Lemma}
\theoremstyle{definition}
\newtheorem{definition}{Definition}
\newtheorem{proposition}{Proposition}
\usepackage[ruled,linesnumbered]{algorithm2e}
\usepackage{enumitem}

\begin{document}

\title{Distributionally Robust Contract Theory for Edge AIGC Services in Teleoperation}

\markboth{Journal of \LaTeX\ Class Files,~Vol.~14, No.~8, August~2015}%
{Shell \MakeLowercase{\textit{et al.}}: Bare Demo of IEEEtran.cls for IEEE Journals}

\author{Zijun~Zhan,
        Yaxian~Dong,
        Daniel~Mawunyo~Doe,
        Yuqing~Hu,
        Shuai~Li,
        Shaohua Cao,
        Lei~Fan,
        and~Zhu~Han,
        \thanks{Zijun Zhan and Lei Fan are with the Department of Electrical and Computer Engineering, University of Houston, 4800 Calhoun Rd, Houston, TX 77004. E-mail: zzhan@uh.edu and lfan8@central.uh.edu} 
         \thanks{Yaxian Dong and Yuqing Hu are with the Department of Architectural Engineering, The Pennsylvania State University, University Park, PA 16802, USA (E-mail: yzd5221@psu.edu and yfh5204@psu.edu)}
	\thanks{Daniel Mawunyo Doe is with the Department of Electrical and Computer Engineering, Prairie View A\&M University, 100 University Dr, Prairie View, TX 77446. Email: dmdoe@pvamu.edu}
         \thanks{Shuai Li is with the Department of Civil \& Coastal Engineering, University of Florida, Gainesville, FL 32611. E-mail: shuai.li@ufl.edu}
         \thanks{Shaohua Cao is with the Qingdao Institute of Software, College of Computer Science and Technology, China University of Petroleum (East China), Qingdao 266580, China. E-mail:shaohuacao@upc.edu.cn}
         \thanks{Zhu Han is with the Department of Electrical and Computer Engineering, University of Houston, 4800 Calhoun Rd, Houston, TX 77004, and also with the Department of Computer Science and Engineering, Kyung Hee University, Seoul, South Korea, 446-701. E-mail: hanzhu22@gmail.com}
}

\maketitle

\begin{abstract}
Advanced AI-Generated Content (AIGC) technologies have injected new impetus into teleoperation, further enhancing its security and efficiency. Edge AIGC networks have been introduced to meet the stringent low-latency requirements of teleoperation. However, the inherent uncertainty of AIGC service quality and the need to incentivize AIGC service providers (ASPs) make the design of a robust incentive mechanism essential. This design is particularly challenging due to both uncertainty and information asymmetry, as teleoperators have limited knowledge of the remaining resource capacities of ASPs. To this end, we propose a distributionally robust optimization (DRO)-based contract theory to design robust reward schemes for AIGC task offloading. Notably, our work extends the contract theory by integrating DRO, addressing the fundamental challenge of contract design under uncertainty. In this paper, contract theory is employed to model the information asymmetry, while DRO is utilized to capture the uncertainty in AIGC service quality. Given the inherent complexity of the original DRO-based contract theory problem, we reformulate it into an equivalent, tractable bi-level optimization problem. To efficiently solve this problem, we develop a Block Coordinate Descent (BCD)-based algorithm to derive robust reward schemes. Simulation results on our unity-based teleoperation platform demonstrate that the proposed method improves teleoperator utility by 2.7\% to 10.74\% under varying degrees of AIGC service quality shifts and increases ASP utility by 60.02\% compared to the SOTA method, i.e., Deep Reinforcement Learning (DRL)-based contract theory. The code and data are publicly available at https://github.com/Zijun0819/DRO-Contract-Theory.
\end{abstract}


\IEEEpeerreviewmaketitle

\section{Introduction} \label{sec:1}
\IEEEPARstart{W}{ith} the maturity of 5G and coming of 6G, virtual reality (VR), augmented reality (AR), and robotics, a new working paradigm—teleoperation—has emerged, offering greater efficiency, enhanced security, and reduced costs compared to traditional working models \cite{darvish2023teleoperation}. As a result, teleoperation has gained widespread attention across various sectors, including unmanned aerial vehicles (UAVs) \cite{javed2024state}, remotely operated vehicles (ROVs) \cite{ata2024intelligent}, construction automation \cite{lee2022challenges}, and medical teleoperation \cite{navarro2022new}. Furthermore, in light of the recent advancements in Artificial Intelligence-Generated Content (AIGC) technologies \cite{cao2025survey}, a plethora of scholars endeavor to unleash the power of AIGC to further advance teleoperation \cite{zhan2024vision, li2024industrial, mu2024robotwin}. In this work, we focus on a representative case involving a Generative Diffusion Model (GDM)-based AIGC empowered ROV operations on construction sites at night \cite{zhan2024vision}. Specifically, the GDM service is capable of enhancing low-light frames or video streams by converting them into high-quality, normal-light outputs. This enables seamless teleoperation in dark environments such as tunnels, nighttime construction sites, and drainage systems.

\begin{figure}[!t]
    \begin{center}
        \includegraphics[width=\linewidth]{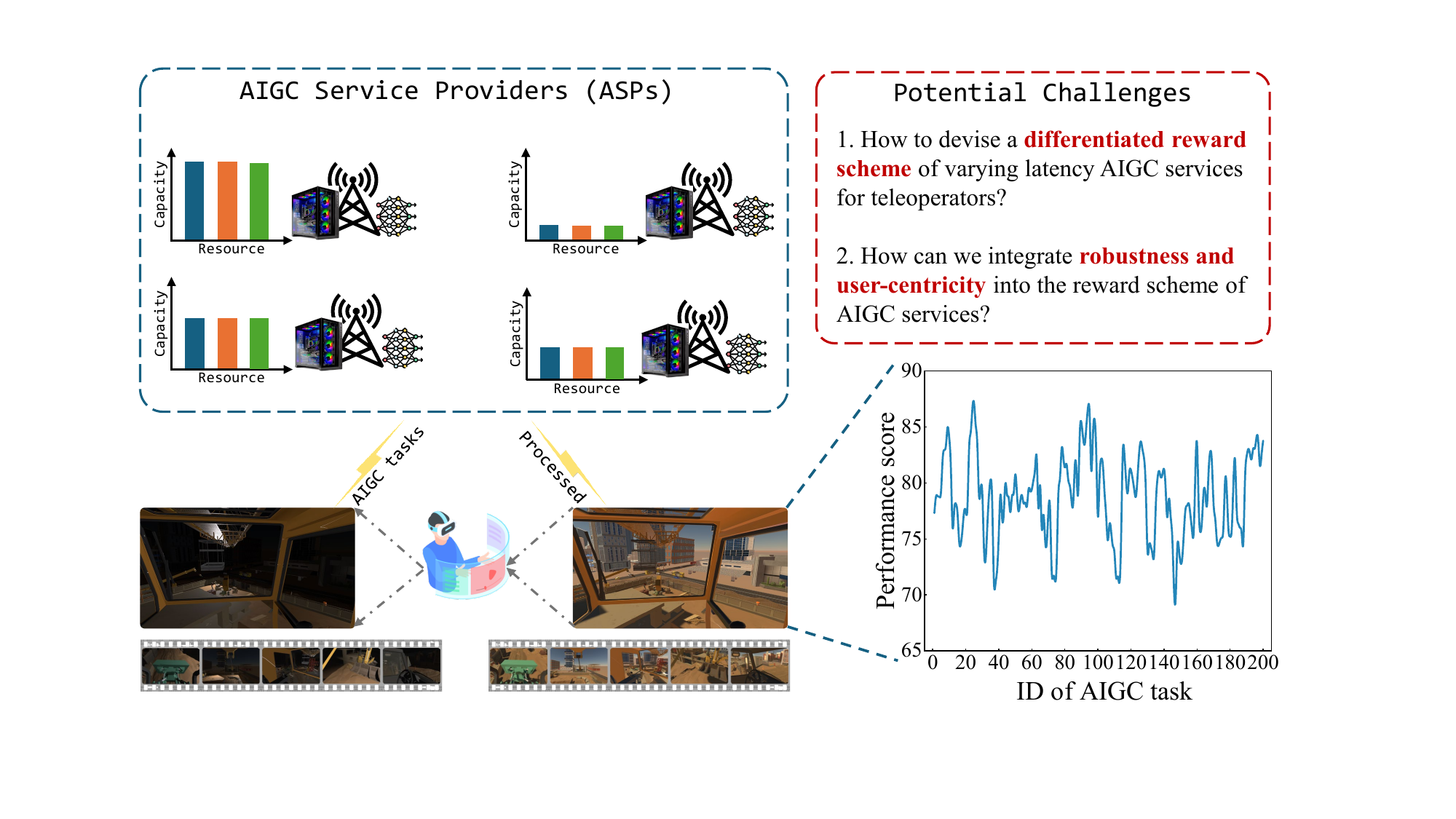}
        \caption{The potential challenges of AIGC-empowered edge teleoperation, where the performance score is acquired via LPIPS \cite{zhang2018unreasonable} and SSIM \cite{wang2004image} of processed AIGC results.}
        \label{fig1}
    \end{center}
\end{figure}

Given the constrained on-board computational resources and the high computational demands of the GDM service, ensuring low-latency teleoperation remains a significant challenge \cite{kamtam2024network}. To overcome this limitation, numerous studies have proposed AIGC service offloading frameworks \cite{du2024diffusion, du2024enabling, li2024multi}, where AIGC tasks are offloaded to AIGC service providers (ASPs) equipped with well-trained AIGC models, abundant computational resources, and physical proximity to the ROVs. While such offloading effectively mitigates the latency bottleneck in GDM-empowered teleoperation, it simultaneously introduces new challenges, as illustrated in Fig. \ref{fig1}. Concretely, the computational resources and queue sizes of different ASPs vary dynamically, necessitating the design of a differentiated reward scheme for teleoperators and ASPs. Moreover, AIGC service quality is inherently uncertain, varying across different tasks even when using the same model, as shown in the sub-figure of Fig. \ref{fig1}. This variability directly impacts the utility of teleoperators, making robust reward scheme design critical. Therefore, we consider designing a robust, differentiated reward scheme from the perspective of the teleoperator, aiming to attract more teleoperators to pay for the AIGC service. Nonetheless, formulating such a reward scheme should account for information asymmetry \cite{li2023book}, as teleoperators have limited knowledge of the real-time computational resources and queue statuses of ASPs. Upon the preceding illustration, we propose a research questions tackled in this paper, which is stated as
\begin{itemize}
    \item[Q1:] How can a robust and differentiated reward scheme be formulated under information asymmetry, so as to optimize the utility of both teleoperators and ASPs?
\end{itemize}

The potential solution to Q1 is contract theory \cite{li2023book}, which excels at formulating differentiated contract bundles, i.e., reward schemes, for varying latencies of AIGC services under information asymmetry. For instance, the authors in \cite{ye2024optimizing} applied contract theory to an AIGC task offloading scenario where the ASP has limited knowledge of users' subjective gains from the AIGC service. Similarly, the authors in \cite{wen2024diffusion} leveraged contract theory to address the issue of users being unaware of the complexity of the AIGC model deployed on ASPs. Furthermore, given that ASPs do not know the difficulty level distribution of AIGC tasks, the authors in \cite{zhan2024vision} used contract theory to design a differentiated pricing strategy for optimizing ASPs' utility under information asymmetry. In addition, many researchers \cite{wen2023freshness, wen2024learning, liu2024prosecutor} have developed effective incentive mechanisms for various types of information asymmetries in AIGC task offloading scenarios using contract theory.

Notably, traditional contract theory can only partially address Q1, as the uncertain quality of AIGC services is difficult to model using contract theory alone. One potential solution is to leverage deep reinforcement learning (DRL)-based contract theory \cite{zhan2023, lotfi2024semantic, zhan2025deep} to fully tackle Q1. Nonetheless, DRL-based contract theory has certain limitations, including high training costs and a lack of robustness in environments that deviate from the training conditions. Alternatively, another promising solution for the uncertainty aspect of Q1 is distributionally robust optimization (DRO) \cite{mohajerin2018data, lin2022distributionally}, an advanced optimization framework that differs from traditional deterministic optimization. DRO excels at handling uncertain optimization problems with relatively low computational cost and strong robustness in shifting environments. Therefore, we propose integrating DRO into contract theory to address Q1 more effectively.

The framework of our proposed DRO-based contract theory is presented in Fig. \ref{fig2}. Specifically, given that the teleoperator only possesses partial information regarding the remaining computational capacity and queue size of the ASP when attempting to offload an AIGC task, we leverage contract theory to model and formulate a series of contract bundles, \{contract type, (reward, latency)\}, aiming to optimize the utilities of both the teleoperator and the ASP. For clarity, we assume that the AIGC model deployed on each ASP is identical. Furthermore, considering the uncertainty in AIGC service quality, we adopt Wasserstein-based DRO approach to model the potential quality distribution of the AIGC service, a.k.a. uncertainty set, which is detailed in (\ref{eq:7}), and optimize the contract bundles against all possible distributions. This allows us to obtain a series of robust contract bundles, \textbf{\{contract type, (reward, quality, latency)\}}. With these robust contract bundles, the teleoperator's utility will not degrade significantly, even when the quality of the AIGC service fluctuates or shifts considerably. In summary, the main contributions of this paper are as follows

\begin{enumerate}
    \item[1.] A framework of DRO-based contract theory for AIGC task offloading in teleoperation is proposed. By employing contract theory, the proposed framework effectively derives reward schemes for teleoperators under information asymmetry. Furthermore, DRO is integrated into the contract theory framework to model the uncertain quality of AIGC services, enhancing the robustness of the contract bundles.

    \item[2.] We extend the classical contract theory problem by integrating DRO, which is capable of tackling contract design under uncertainty. Concretely, we formulate the DRO-based contract theory problem and theoretically prove that the original intractable problem can be equivalently reformulated as a tractable bi-level optimization problem.

    \item[3.] Given the bi-level structure of the reformulated optimization problem, we propose a Block Coordinate Descent (BCD)-based contract theory algorithm to solve the DRO-based contract theory problem. We apply the BCD algorithm to iteratively solve the optimization problem and obtain the robust contract bundles. Additionally, we demonstrate the convergence of our proposed algorithm through experimental results.
    
    \item[4.] Numerical simulation experiments on a unity-based teleoperation project are conducted. The results demonstrate that the proposed framework effectively optimizes the utilities of both teleoperators and ASPs. Compared to DRL-based contract theory, the proposed algorithm improves the average utility of teleoperators by $2.7 \sim 10.74\%$ across various AIGC service quality shift scenarios and augments ASP utility by $60.02\%$.  
\end{enumerate}

The rest of this paper is structured as follows. Section \ref{sec:2} provides a systematic literature review. Section \ref{sec:3} presents the proposed DRO-based contract theory framework, along with relevant preliminary knowledge and the system model. In Section \ref{sec:4}, we formulate the DRO-based contract theory problem and equivalently transform the complex original problem into a tractable bi-level optimization problem. Section \ref{sec:5} elaborates on and analyzes the proposed BCD-driven contract theory algorithm. Section \ref{sec:6} systematically evaluates the proposed framework. Finally, Section \ref{sec:7} concludes the paper.

\section{Related Works} \label{sec:2}
In this section, we review several topics that are related to our work, which are contract theory in edge AIGC networks in Section \ref{sec:2.1} and uncertain optimization in Section \ref{sec:2.2}, respectively.

\subsection{Contract Theory in Edge AIGC Networks} \label{sec:2.1}
With the development of various AIGC techniques, numerous researchers have sought to expand the impact of AIGC through edge AIGC \cite{du2024diffusion, du2024enabling, li2024multi}, enabling local devices to access AIGC services in real-time. Information asymmetry is prevalent in edge AIGC networks, such as cases where the complexity of the AIGC model is unknown to users or the difficulty level distribution of AIGC tasks is private information held by the ASP, which hinders the design of effective incentive mechanisms for edge AIGC services. To address this, many researchers have contributed to incentive mechanism design in edge AIGC networks using contract theory.

For example, the authors in \cite{ye2024optimizing} proposed a two-stage incentive mechanism based on contract theory for AIGC task allocation, where the ASP does not know the users' subjective gain per unit of service quality. The authors in \cite{wen2024diffusion} designed a user-centric incentive mechanism using contract theory and prospect theory, addressing information asymmetry where users are unaware of the AIGC model's complexity. Similarly, the authors in \cite{zhan2024vision} developed a vision-language model-empowered contract theory approach in scenarios where the ASP possesses only partial information about the difficulty level distribution of AIGC tasks. Additionally, the authors in \cite{wen2023freshness, wen2024learning} proposed age-of-information-based contract theory mechanisms to incentivize UAVs or mobile devices to collect fresh data for AIGC model fine-tuning, where data collection costs are private information. Moreover, the authors in \cite{liu2024prosecutor} introduced contract theory into blockchain-based AIGC networks, where users lack knowledge of the resources invested by ASPs, and contract theory is applied to motivate ASPs to devote more resources for higher profits.

These studies explore the integration of contract theory into edge AIGC networks from various perspectives and lay the foundation for future research. However, ensuring the robustness of contract design remains a challenge due to uncertainty in AIGC service quality. To address this, our study enhances contract theory robustness by integrating DRO into edge AIGC networks.

\subsection{Optimization under Uncertainty} \label{sec:2.2}
Given the uncertainty of AIGC service quality posing a challenge for robust contract bundle design, this section reviews several uncertainty optimization approaches.

Firstly, considering that uncertainty often implies dynamic changes to some extent, and DRL is well-suited for adapting to dynamic environments, some studies have explored leveraging DRL to obtain contract bundles \cite{zhan2023, lotfi2024semantic, zhan2025deep}. Notably, DRL-based contract theory can derive contract bundles even when the contract theory problem is high-dimensional and non-convex. However, the performance of DRL-based contract theory degrades when the environment undergoes significant shifts.

Secondly, in contrast to deep learning-based DRL methods, another line of research addresses uncertainty from the traditional optimization perspective, including stochastic programming (SP) \cite{schneider2006stochastic}, robust optimization (RO) \cite{bertsimas2011theory}, and DRO \cite{kuhn2024}. Among these, DRO has been shown to outperform SP and RO in terms of robustness and performance under uncertainty \cite{lin2022distributionally}. As a result, numerous studies have applied DRO to solve various uncertain optimization problems. For example, the authors in \cite{yin2024distributionally, xia2022preserving, wang2023multi} employed DRO to address uncertainty in renewable energy systems, while the authors in \cite{wu2023understanding, jiang2022learning} explored the application of DRO in deep learning to enhance model robustness.

Distinct from previous works, this paper aims to integrate DRO into contract theory to derive more robust contract bundles for AIGC service offloading. To the best of our knowledge, this is the first work to combine DRO with contract theory for robust incentive mechanism design.

\section{Preliminary Knowledge and System Model} \label{sec:3}
In this section, we first briefly review the framework of our proposed DRO-based contract theory in Section \ref{sec:3.1}. Then, we introduce the preliminary knowledge relevant to this paper in Section \ref{sec:3.2}. Finally, the utility models of teleoperators and ASPs are formulated in Section \ref{sec:3.3}.

\begin{figure}[!t]
    \begin{center}
        \includegraphics[width=\linewidth]{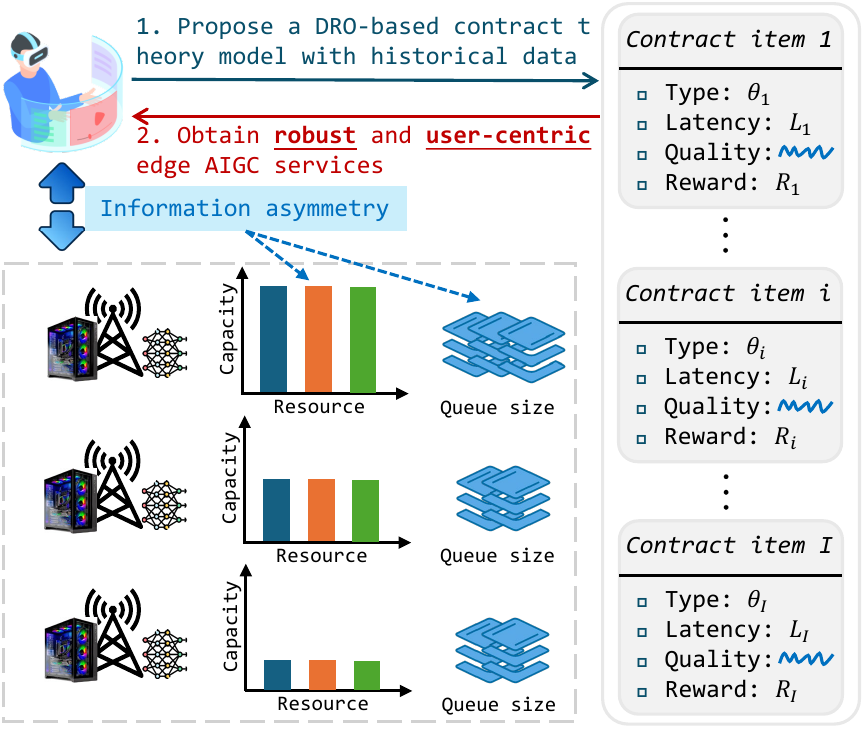}
        \caption{Framework of the distributional robust optimization-based contract theory.}
        \label{fig2}
    \end{center}
\end{figure}

\subsection{Framework Overview} \label{sec:3.1}
In this paper, we consider the scenario of teleoperators working at night who require edge AIGC services to efficiently process low-light images (AIGC tasks) into normal-light images, so as to ensure secure operation. As depicted in Fig. \ref{fig2}, different edge ASPs may have varying resource capacities and queue sizes, creating the need to formulate a differentiated pricing strategy for AIGC services with varying latencies. Moreover, considering the importance of user-centricity for attracting more teleoperators to pay for the AIGC service, we assume that the differentiated pricing strategy is designed from the perspective of teleoperators rather than ASPs. However, one key challenge is information asymmetry, where teleoperators have only partial knowledge of the resource capacities and queue sizes of ASPs. Additionally, a second challenge arises from the uncertainty of AIGC service quality, which can be viewed from two perspectives. First, the subjective utility of different teleoperators for the same AIGC output may vary. For instance, some users might prefer sharp images, while others may favor softer, more natural-looking visuals, artistic styles, or enhanced contrast for better visibility in low-light conditions. Second, the objective utility of teleoperators differs across various AIGC tasks. For example, as depicted in the subfigure of Fig. \ref{fig1}, the objective performance scores of different AIGC tasks vary.

To address these two challenges, as illustrated in Fig. \ref{fig2}, we propose a Wasserstein-driven DRO-based contract theory framework for differentiated pricing scheme formulation. In this framework, contract theory is employed to tackle the challenge of information asymmetry, while Wasserstein-based DRO is integrated into the contract theory to address the uncertainty of AIGC service quality. Based on this approach, a series of user-centric and robust contract bundles can be formulated for AIGC task offloading, thereby attracting more teleoperators to pay for the edge AIGC services.

\subsection{Preliminary Knowledge} \label{sec:3.2}
\subsubsection{Conditional Diffusion Model} \label{sec:3.2.1}
The AIGC service utilized in this paper is a conditional generative diffusion-based AIGC model, which consists of two phases: the forward process and the reverse process \cite{ho2020denoising} \cite{song2021denoising}.

The forward process is a diffusion process in which noise is gradually added to the original data until it becomes pure noise that follows a Gaussian distribution. Mathematically, the final pure noise $\mathbf{x}_T \sim \mathcal{N}(0, \mathbf{I})$ is progressively transformed from the original data $\mathbf{x}_0$ via
\begin{equation} \label{eq:1}
    q (\mathbf{x}_{T}|\mathbf{x}_0) = \prod_{t=1}^T q(\mathbf{x}_{t}|\mathbf{x}_{t-1})
\end{equation}
and
\begin{equation} \label{eq:2}
    q(\mathbf{x}_{t}|\mathbf{x}_{t-1})=\mathcal{N}(\mathbf{x}_t;\sqrt{1-\beta_t}\mathbf{x}_{t-1}, \beta_t\mathbf{I})
\end{equation}
 in $T$ steps. Here, $\beta_t$ is the schedule parameter of the diffusion model.

 The reverse process, a.k.a. denoising process, aims to deduce the target $\hat{\mathbf{x}}_0$ from the noise sample $\hat{\mathbf{x}}_T \sim \mathcal{N}(0, \mathbf{I})$ and conditional input $\tilde{x}$, i.e.,
 \begin{equation} \label{eq:3}
    p_{\phi}(\hat{\mathbf{x}}_0) = p(\hat{\mathbf{x}}_T) \prod_{t=1}^T p_{\phi}(\hat{\mathbf{x}}_{t-1}|\hat{\mathbf{x}}_t, \tilde{\mathbf{x}}),
\end{equation}
 Here, $p_{\phi}(\hat{\mathbf{x}}_{t-1}|\hat{\mathbf{x}}_t, \tilde{x})$ is akin to (\ref{eq:2}),
 \begin{equation} \label{eq:4}
     p_{\phi}(\hat{\mathbf{x}}_{t-1}|\hat{\mathbf{x}}_t, \tilde{\mathbf{x}})=\mathcal{N}(\tilde{\mathbf{x}}_{t-1};\mathbf{\mu_\phi}(\hat{\mathbf{x}}_{t}, t, \tilde{\mathbf{x}}), \sigma^2_t\mathbf{I}),
 \end{equation}
 where $\mathbf{\mu_\phi}(\hat{\mathbf{x}}_{t}, t, \tilde{\mathbf{x}})$ and $\sigma^2_t$ are approximated value of $\tilde{\mathbf{x}}_{t-1}$ and variance parameter, respectively. Here, $\mathbf{\mu_\phi}(\hat{\mathbf{x}}_{t}, t, \tilde{\mathbf{x}})$ is defined as
\begin{equation} \label{eq:5}
    \mathbf{\mu_\phi}(\hat{\mathbf{x}}_{t}, t, \tilde{\mathbf{x}})=\frac{1}{\sqrt{\alpha_t}}(\hat{\mathbf{x}}_{t} - \frac{\beta_t}{\sqrt{1 - \bar{\alpha}_t}} \epsilon_\phi(\hat{\mathbf{x}}_{t}, t, \tilde{\mathbf{x}})),
\end{equation}
where $\alpha_t = 1 - \beta_t, \bar{\alpha}_t=\prod_{1}^{t} \alpha_t$, and $\epsilon_\phi(\hat{\mathbf{x}}_{t}, t, \tilde{\mathbf{x}})$ represents a neural network for noise, adds in the forward process, prediction.

\subsubsection{Wasserstein-Based Distributional Robust Optimization} \label{sec:3.2.2}
To safeguard the utility of the teleoperator under the uncertainty of AIGC services, we adopt Wasserstein-based DRO \cite{kuhn2019wasserstein} to assist in contract bundle design. The fundamental rationale of Wasserstein-based DRO is to approximate a set of potential AIGC service quality distributions and then optimize over this set. Specifically, the teleoperator constructs an empirical distribution $\tilde{\mathbb{P}}$ of AIGC service quality based on collected historical samples $\{\tilde\xi_n | n \in [N]\}$, where $[N]=\{1, \cdots, N\}$. The empirical distribution $\tilde{\mathbb{P}}$ is defined as
\begin{equation} \label{eq:6}
    \tilde{\mathbb{P}} = \frac{1}{N}\sum_{n=1}^{N}{\delta_{\tilde\xi_n}},
\end{equation}
where $\delta_{\tilde\xi_n}$ denotes the Dirac measure of $\tilde\xi_n$.

Since the empirical distribution $\tilde{\mathbb{P}}$ may be inaccurate and may not reflect the true distribution $\hat{\mathbb{P}}$ of AIGC service quality, we construct an ambiguity set $\tilde{\mathcal{P}}$ around $\tilde{\mathbb{P}}$, which, with high confidence level $\tau$, contains the true distribution $\hat{\mathbb{P}}$. The ambiguity (uncertainty) set $\tilde{\mathcal{P}}$ is defined as
 \begin{equation} \label{eq:7}
     \tilde{\mathcal{P}} = \left\{\mathbb{P} \Big| W\left(\tilde{\mathbb{P}},\mathbb{P}\right) \leq \epsilon(N)\right\},
 \end{equation}
where $W\left(\tilde{\mathbb{P}},\mathbb{P}\right)$ denotes the Wasserstein distance between $\tilde{\mathbb{P}}$ and $\mathbb{P}$, calculated as
\begin{equation} \label{eq:8}
    W\left(\tilde{\mathbb{P}},\mathbb{P}\right) = \inf_{\Pi} \left\{ \int_{\Xi^2} \|\tilde\xi - \xi\| \,  \Pi(d\tilde\xi, d\xi) \right\}.
\end{equation}
Here, $\Pi$ represents the joint distribution of $\tilde\xi$ and $\xi$, and $\Xi$ denotes the support set of the AIGC service quality. It is worth noting that the $\|\cdot\|_1$ norm is adopted in this paper to ensure the numerical tractability of the DRO \cite{yin2024distributionally, xia2022preserving, wang2022distributionally}.

The radius of the ambiguity set $\epsilon(N)$ in (\ref{eq:7}) controls the size of $\tilde{\mathcal{P}}$ and is defined as
\begin{equation} \label{eq:9}
    \epsilon(N) = D \sqrt{\frac{2}{N} \ln \left( \frac{1}{1 - \tau} \right)},
\end{equation}
where $D$ is the diameter of $\Xi$ \cite{duan2018distributionally}.

\subsection{Utility Model of the ASP and Teleoperator} \label{sec:3.3}
As described above, we assume there are $M$ edge ASPs in the edge AIGC system. For simplicity, we assume that each ASP is equipped with an identical AIGC model and can be categorized into $I$ types, denoted as ${\theta_i | i \in [I]}$, based on their resource capacities and queue sizes. Here, $\theta_i$ represents the willingness of a type-$i$ ASP to process AIGC tasks, which is proportional to the resource capacity and queue size of the ASP. Furthermore, the monotonicity condition holds for $\theta_i$ and is expressed as
\begin{equation} \label{eq:10}
    0 < \theta_1 \leq \cdots \leq \theta_i \leq \cdots \leq \theta_I.
\end{equation}

Analogous to \cite{wen2024diffusion, liu2024deep}, we consider that the utility of a type-$i$ ASP primarily depends on the difference between the reward received for processing an AIGC task and the associated resource cost. Accordingly, the utility of a type-$i$ ASP is defined as
\begin{equation} \label{eq:11}
    \pi _A^i = {\theta _i}{R_i} - {\gamma _1}{L_i},
\end{equation}
where $L_i$ is the inverse of the latency requirement. For example, if $L_i = 20$, the corresponding latency requirement is $\frac{1}{20}$ seconds, i.e., $50$ ms. $R_i$ is the reward the teleoperator needs to pay the ASP if the AIGC service is delivered in the latency requirement of $1/L_i$ ms. $\gamma_1$ is the coefficient used to map the latency to the resource cost, which is greater than 0.

Analogously, we consider that the utility of the teleoperator is the difference between the benefit gained from receiving AIGC services within the latency requirement and the reward paid to the ASP. Moreover, the benefit of the teleoperator is proportional to the quality and latency of the AIGC service. Therefore, the utility of the teleoperator under a type-$i$ ASP is formulated as
\begin{equation} \label{eq:12}
    \pi _T^i = \ln ({\gamma _2}\xi  + {\gamma _3}{L_i}) - {R_i}.
\end{equation}
Here, $\xi$ represents the uncertain subjective utility of the teleoperator regarding the AIGC service provided by the type-$i$ ASP. $\gamma_2$ and $\gamma_3$ are coefficients used to balance the weights of the quality and latency of the AIGC service, respectively.

\section{Problem Statement and Reformulation} \label{sec:4}
Based on the system model presented in Section \ref{sec:3}, we first formulate the DRO-based contract theory problem in Section \ref{sec:4.1} and then reformulate the problem into a tractable form in Section \ref{sec:4.2}.

\subsection{Problem Statement} \label{sec:4.1}
Based on the preceding sections, the objective of this paper is to optimize the teleoperator's utility under uncertain AIGC service quality and information asymmetry. To address information asymmetry, we adopt contract theory to formulate the optimization problem. The basic rationale of contract theory is to optimize the teleoperator's utility through a series of carefully constructed contract bundles, $\{(L_i, R_i) | i \in [I]\}$. Notably, the contract bundles must simultaneously satisfy the constraints of Incentive Rationality (IR) and Incentive Compatibility (IC) to ensure that each type of ASP selects the contract bundle specifically designed for them.
\begin{definition}[\textit{Incentive Rationality}] \label{definition1}
	The finalized $\{(L_i, R_i) | i \in [I]\}$ should guarantee that the utility of each type of ASP is greater than 0,i.e.,
    \begin{equation} \label{eq:13}
        \pi _A^i(L_i, R_i) \geq 0, \forall i \in [I].
    \end{equation}
\end{definition}
Here, the IR constraint ensures that the ASP is willing to participate in AIGC task offloading. To safeguard the utility of the teleoperator \cite{zhan2023}, it is also necessary to ensure that the IC constraint is satisfied.
\begin{definition}[\textit{Incentive Compatibility}] \label{definition2}
	The highest utility of the type-$i$ ASP is achieved when they select the corresponding contract bundle $(L_i, R_i)$, which is defined as
    \begin{equation} \label{eq:14}
        \pi _A^i(L_i, R_i) \geq \pi _A^i(L_j, R_j), \forall i,j \in [I], j \neq i.
    \end{equation}
\end{definition}
If the IC constraint is not satisfied, the type-$i$ ASP may choose other contract bundles, causing the intended teleoperator utility to fail and ultimately impairing the teleoperator's utility.

Notably, the teleoperator's utility is coupled with the uncertainty of AIGC service quality, and this uncertainty must be considered when formulating the optimization problem using contract theory. In this paper, we employ Wasserstein-based DRO to model and address this uncertainty within the contract theory framework. Accordingly, the optimization problem is formulated as
\begin{subequations} \label{eq:15}
	\begin{align}
		\sup_{\mathbf{L}, \mathbf{R}} \inf_{\mathbb{P} \in \tilde{\mathcal{P}}} ~ 
&\mathbb{E}^{\mathbb{P}} \left[ \sum_i \alpha_i \pi_{T}^i \right]\label{Problem_1} \\
		  s.t. & ~ (\ref{eq:13}), (\ref{eq:14}), \label{P1_constraint_1}
	\end{align}
\end{subequations}
where $\alpha_i$ denotes the probability of a type-$i$ ASP existing in the AIGC offloading system, where $\sum_{i=1}^{I}{\alpha_i} = 1$. Eq. (\ref{eq:13}) includes $I$ IR constraints, and Eq. (\ref{eq:14}) includes $I(I-1)$ IC constraints. The physical meaning of (\ref{Problem_1}) is to find a robust solution for the contract bundles that accounts for the worst-case distribution of AIGC service quality, ensuring that the teleoperator's utility remains resilient under varying conditions.

\subsection{Problem Reformulation} \label{sec:4.2}
Given (\ref{Problem_1}) is a maxmin problem and $I^2$ constraints exist, rendering (\ref{Problem_1}) numerically intractable. To this end, in this section, we endeavor to reformulate (\ref{Problem_1}) in a tractable form.

Initially, we sought to reduce the constraints in (\ref{Problem_1}) and obtained the following proposition.
\begin{proposition} \label{proposition_1}
\textit{With constraints deduction, the problem in (\ref{Problem_1}) can be transformed into}
    \begin{subequations} \label{eq:16}
	\begin{align}
		\sup_{\mathbf{L}} \inf_{\mathbb{P} \in \tilde{\mathcal{P}}} ~ 
&\mathbb{E}^{\mathbb{P}} \left[ \sum_{i=1}^{I} \alpha_i \pi_{T}^{i} \right]   \label{Problem_2} \\
		  s.t. & ~ R_i = \gamma_1\left(\frac{ L_1}{\theta_1} + \sum_{j=2}^{i} \frac{L_j - L_{j-1}}{\theta_j}\right), \label{P2_constraint_1} \\
          & ~ L_1 \leq \dots \leq L_i \leq \dots \leq L_I. \label{P2_constraint_2}
	\end{align}
\end{subequations}

\begin{proof}
    The proof can be proved via following Lemmas.
    \begin{lemma} \label{lemma_1}
        \textbf{(Monotonicity constraint)} $\theta_i \geq \theta_j \Leftrightarrow R_i \geq R_j \Leftrightarrow L_i \geq L_j$ hold for $\forall i, j \in [I]$.
        \begin{proof}
            With IC constraints, we can obtain
            \begin{equation} \label{eq:17}
                \theta_i R_i - \gamma_1 L_i \geq \theta_i R_j - \gamma_1 L_j
            \end{equation}
            and
            \begin{equation} \label{eq:18}
                \theta_j R_j - \gamma_1 L_j \geq \theta_j R_i - \gamma_1 L_i
            \end{equation}
            hold for $\forall i, j \in [I]$.
            
            With (\ref{eq:17}) and (\ref{eq:18}), we can deduce
            \begin{equation} \label{eq:19}
                \begin{aligned}
                    &(\theta_i - \theta_j) (R_i - R_j)  > 0, \\
                    &\gamma_1 (L_i - L_j) \geq \theta_j (R_i - R_j),
                \end{aligned}
            \end{equation}
            which indicates $\theta_i \geq \theta_j \Leftrightarrow R_i \geq R_j \Leftrightarrow L_i \geq L_j$ since $\gamma_1 > 0$ holds. Therefore, the proof of Lemma \ref{lemma_1} is completed.
        \end{proof}
    \end{lemma}

    With Lemma \ref{lemma_1}, the monotonicity of $\theta$ in (\ref{eq:10}) can be extended to $R$ and $L$, i.e.,
    \begin{equation} \label{eq:20}
        R_1 \leq \cdots \leq R_i \leq \cdots \leq R_I
    \end{equation}
    and
    \begin{equation} \label{eq:21}
        L_1 \leq \cdots \leq L_i \leq \cdots \leq L_I.
    \end{equation}

    \begin{lemma} \label{lemma_2}
        \textbf{(IR constraint reduction)} The $I$ IR constraints can be reduced to $\pi _A^{1}({L_{1}},{R_{1}}) = 0$ if (\ref{eq:10}) is satisfied.
        \begin{proof}
            With IC and monotonicity constraints, we can derive the following inequalities
            \begin{equation} \label{eq:22}
                \theta_i R_i - \gamma_1 L_i \geq \theta_i R_1 - \gamma_1 L_1 \geq \theta_1 R_1 - \gamma_1 L_1, \forall i \in [I].
            \end{equation}
            With (\ref{eq:22}), the proof of Lemma \ref{lemma_2} is done.
        \end{proof}
    \end{lemma}

    \begin{lemma} \label{lemma_3}
        \textbf{(IC constraint reduction)} The $I(I-1)$ IC constraints can be reduced to $\pi _A^{i+1}({L_{i+1}},{R_{i+1}}) = \pi _A^{i+1}({L_{i}},{R_{i}}),\forall i \in [I-1] $ if (\ref{eq:10}) is satisfied.
        \begin{proof}
            With IC and monotonicity constraints, following inequalities for $\forall i \in [I-1]$ can be derived
            \begin{equation} \label{eq:23}
                \theta_{i+1} R_{i+1} - \gamma_1 L_{i+1} \geq \theta_{i+1} R_{i} - \gamma_1 L_{i}, 
            \end{equation}
            \begin{equation} \label{eq:24}
                \theta_{i} R_{i} - \gamma_1 L_{i} \geq \theta_{i} R_{i-1} - \gamma_1 L_{i-1},
            \end{equation}
            and
            \begin{equation} \label{eq:25}
                \theta_{i+1} (R_{i} - R_{i-1}) \geq \theta_{i} (R_{i} - R_{i-1}).
            \end{equation}
            Subsequently, adding (\ref{eq:23}), (\ref{eq:24}), and (\ref{eq:25}) together, we can derive
            \begin{equation} \label{eq:26}
                \theta_{i+1} R_{i+1} - \gamma_1 L_{i+1} \geq \theta_{i+1} R_{i-2} - \gamma_1 L_{i-2}.
            \end{equation}
            Repeating the similar process of (\ref{eq:23})-(\ref{eq:26}), we can formulate the following inequalities hold for $\forall i \in [I-1]$,
            \begin{equation} \label{eq:27}
                \begin{aligned}
                    \theta_{i+1} R_{i+1} - \gamma_1 L_{i+1} &\geq \theta_{i+1} R_{i-2} - \gamma_1 L_{i-2} \\
                    &\cdots \\
                    \theta_{i+1} R_{i+1} - \gamma_1 L_{i+1} &\geq \theta_{i+1} R_{1} - \gamma_1 L_{1}.
                \end{aligned}
            \end{equation}
            Here, Eq. (\ref{eq:27}) indicates if the local downward IC constraint is satisfied, all the downward IC constraints are inherently satisfied.

            Analogously, taking similar deriving process of (\ref{eq:23})-(\ref{eq:27}), we can deduce that if the local upward IC constraint is satisfied, all the upward IC constraints are inherently satisfied for $\forall i \in [I-1]$, i.e.,
            \begin{equation} \label{eq:28}
                \begin{aligned}
                    \theta_{i} R_{i} - \gamma_1 L_{i} &\geq \theta_{i} R_{i+3} - \gamma_1 L_{i+3} \\
                    &\cdots \\
                    \theta_{i} R_{i} - \gamma_1 L_{i} &\geq \theta_{i} R_{I} - \gamma_1 L_{I}.
                \end{aligned}
            \end{equation}

            Additionally, if the local downward IC constraint is binding, the local upward IC constraint will satisfied automatically. Concretely, with
            \begin{equation} \label{eq:29}
                \theta_{i+1} R_{i+1} - \gamma_1 L_{i+1} = \theta_{i+1} R_{i} - \gamma_1 L_{i}
            \end{equation}
            and the monotonicity constraint
            \begin{equation} \label{eq:30}
                \theta_{i+1}(R_{i+1} - R_{i}) \geq \theta_{i}(R_{i+1} - R_{i}),
            \end{equation}
            we can derive
            \begin{equation} \label{eq:31}
                \theta_{i} R_{i} - \gamma_1 L_{i} \geq \theta_{i} R_{i+1} - \gamma_1 L_{i+1}
            \end{equation}
            hold for $\forall i \in [I-1]$.

            In summary,  the proof of Lemma \ref{lemma_3} is completed via (\ref{eq:27})-(\ref{eq:31}).
        \end{proof}
    \end{lemma}

    With Lemmas (\ref{lemma_1}), (\ref{lemma_2}) and (\ref{lemma_3}), the problem in (\ref{eq:15}) can be reformulated as
    \begin{subequations}
        \begin{align}
            \sup_{\mathbf{L}, \mathbf{R}} \inf_{\mathbb{P} \in \tilde{\mathcal{P}}} \quad 
    &\mathbb{E}^{\mathbb{P}} \left[ \sum_i \alpha_i \pi_{T}^i \right]\label{Problem_2_1} \\
              s.t. & \quad \pi _A^{1}({L_{1}},{R_{1}}) = 0, \label{P2_1_constraint_1} \\
                    & \quad \pi _A^{i+1}({L_{i+1}},{R_{i+1}}) = \pi _A^{i+1}({L_{i}},{R_{i}}), \label{P2_1_constraint_2}  \\
                    & \quad L_1 \leq \cdots \leq L_i \leq \cdots \leq L_I. \label{P2_1_constraint_3}
        \end{align}
    \end{subequations}
    By resolving (\ref{P2_1_constraint_1}) and (\ref{P2_1_constraint_2}), we can derive
    \begin{equation} \label{eq:33}
         R_i = \gamma_1\left(\frac{ L_1}{\theta_1} + \sum_{j=2}^{i} \frac{L_j - L_{j-1}}{\theta_j}\right), ~ \forall ~ i \in [I].
    \end{equation}

    Therefore, the proof of Proposition \ref{proposition_1} is completed.
    
\end{proof}

\end{proposition}

Secondly, inspired by \cite{kuhn2024}, we intend to further simplify (\ref{eq:16}) by reformulating (\ref{Problem_2}) into a tractable bilevel optimization problem, i.e., Proposition \ref{proposition_2}.

\begin{proposition} \label{proposition_2}
\textit{The problem in (\ref{Problem_2}) is equivalent to}
    \begin{subequations} \label{eq:34}
	\begin{align}
		\sup_{\mathbf{L}, \lambda, \mathbf{s}} & \quad -\lambda \epsilon + \frac{1}{N} \sum_{n=1}^{N} s_n   \label{Problem_3} \\
		  s.t. & \quad \inf_{\xi} \left( \sum_{i=1}^{I} \alpha_i 
\pi_{T}^i + \lambda \| \xi - \tilde{\xi}_n \| \right) \geq s_n, \label{P3_constraint_1} \\
            & \quad R_i = \gamma_1\left(\frac{ L_1}{\theta_1} + \sum_{j=2}^{i} \frac{L_j - L_{j-1}}{\theta_j}\right), \label{P3_constraint_2} \\
            & \quad \xi \in [\underline{\xi}, \bar{\xi}],  \label{P3_constraint_3}\\
          & \quad L_1 \leq \dots \leq L_i \leq \dots \leq L_I, \label{P3_constraint_4} \\
          & \quad \lambda \geq 0. \label{P3_constraint_5}
	\end{align}
\end{subequations}

\begin{proof}
    Firstly, we intend to transform the inner inf problem of (\ref{eq:16}) into a sup problem. As per the definition of $\tilde{\mathcal{P}}$, i.e., (\ref{eq:7}), the inner inf problem of (\ref{eq:16}) is equivalent to
    \begin{subequations} \label{eq:35}
        \begin{align}
            \inf_{\mathbb{P}} & ~ \int_{\Xi}{\left[ \sum_{i=1}^{I} \alpha_i 
\pi_{T}^i \right]} \mathbb{P}(d\xi)   \label{Problem_3_1} \\
		  s.t. & ~ \int_{\Xi^2} \|\tilde\xi - \xi\| \,  \Pi(d\tilde\xi, d\xi) \leq \epsilon. \label{P3_1_constraint_1}
        \end{align}
    \end{subequations}

    Notably, $\tilde{\xi}$ is sampled historical data, discrete value. Hence, problem (\ref{Problem_3_1}) can be further equivalently transform into
    \begin{subequations} \label{eq:36}
        \begin{align}
            \inf_{\mathbb{P}_i} & ~ \sum_{n=1}^{N} \frac{1}{N} \int_{\Xi}{\left[ \sum_{i=1}^{I} \alpha_i \pi_{T}^i \right]} \mathbb{P}_n(d\xi)   \label{Problem_3_2} \\
		  s.t. & ~ \sum_{n=1}^{N} \frac{1}{N} \int_{\Xi} \|\tilde\xi_n - \xi\| \,  \mathbb{P}_n(d\xi) \leq \epsilon. \label{P3_2_constraint_1}
        \end{align}
    \end{subequations}
    Here, $\Pi(d\tilde\xi, d\xi) = \frac{1}{N}\mathbb{P}_n(d\xi)$ and $\mathbb{P}(d\xi) = \int_{\Xi} \frac{1}{N} \mathbb{P}_n(d\xi) = \sum_{i=1}^{N} \frac{1}{N}\mathbb{P}_n(d\xi)$.

    Subsequently, with the utilization of Lagrangian dual method, the problem in (\ref{eq:36}) can be derived as
    \begin{equation} \label{eq:37}
        \begin{aligned}
            \sup_{\lambda \geq 0} \inf_{\mathbb{P}_i} ~ \sum_{n=1}^{N} \frac{1}{N} & \int_{\Xi}{\left[ \sum_{i=1}^{I} \alpha_i \pi_{T}^i \right]} \mathbb{P}_n(d\xi) \\ &+ \lambda\sum_{n=1}^{N} \frac{1}{N}(\int_{\Xi} \|\tilde\xi_n - \xi\| \,  \mathbb{P}_n(d\xi) - \epsilon)
        \end{aligned}
    \end{equation}
    
    By combining the integral component of (\ref{eq:37}), we can deduce (\ref{eq:37}) is equivalent to
    \begin{equation} \label{eq:38}
        \sup_{\lambda \geq 0} -\lambda \epsilon + \inf_{\xi \in \Xi}{\frac{1}{N} \sum_{n=1}^{N}\left[\sum_{i=1}^{I} \alpha_i \pi_{T}^i + \lambda \|\tilde\xi_n - \xi\| \right]},
    \end{equation}
    where the infimum over $\mathbb{P}_i$ can be achieved via the definition of the Dirac distribution.

    With hypograph reformulation, we can reformulate (\ref{eq:38}) as
    \begin{subequations} \label{eq:39}
        \begin{align}
		\sup_{\lambda, \mathbf{s}} & \quad \lambda \epsilon + \frac{1}{N} \sum_{n=1}^{N} s_n   \label{Problem_3_3} \\
		  s.t. & \quad \inf_{\xi \in \Xi} \left( \sum_{i=1}^{I} \alpha_i 
\pi_{T}^i + \lambda \| \xi - \hat{\xi}_n \| \right) \geq s_n, n \in [N] \label{P3_3_constraint_1} \\
          & \quad \lambda \geq 0. \label{P3_3_constraint_2}
	\end{align}
    \end{subequations}
    Here, $\Xi$ is defined as $[\underline{\xi}, \bar{\xi}]$.

    By substituting (\ref{eq:39}) into (\ref{eq:16}), the proof of Proposition \ref{proposition_2} is completed.
\end{proof}
\end{proposition}

\section{Proposed Approach} \label{sec:5}
With Propositions \ref{proposition_1} and \ref{proposition_2}, the original problem (\ref{eq:15}) is more tractable. Therefore, we present the solution of (\ref{eq:15}) in Section \ref{sec:5.1}. Subsequently, we pinpoint the algorithm of BCD-driven contract theory and analyze its complexity in Section \ref{sec:5.2}.

\subsection{Block Coordinate Descent-Driven Solution and Algorithm} \label{sec:5.1}
Prior to tackle (\ref{eq:34}), we need to solve the inner optimization problem of (\ref{P3_constraint_1}) under the constraint of (\ref{P3_constraint_3}). To this end, we start with expanding and organizing (\ref{P3_constraint_1}) and deriving
\begin{equation} \label{eq:40}
    \begin{aligned}
        \inf_{\xi \in [\underline{\xi}, \bar{\xi}]} & \left( \sum_{i=1}^{I} \alpha_i \pi_{T}^i + \lambda \| \xi - \hat{\xi}_n \| \right) \\
        & = \left(\min_{\xi \in [\underline{\xi}, \bar{\xi}]} f_n(\xi;\mathbf{L}, \lambda) \right) - g(\mathbf{L}).
    \end{aligned}
\end{equation}
Here, $f_n(\xi;\mathbf{L}, \lambda)$ is defined as
\begin{equation} \label{eq:41}
    f_n(\xi;\mathbf{L}, \lambda) = \sum_{i=1}^{I} \alpha_i \ln(\gamma_2\xi + \gamma_3L_i) + \lambda \| \xi - \hat{\xi}_n \|
\end{equation}
and $g(\mathbf{L})$ is defined as
\begin{equation} \label{eq:42}
    g(\mathbf{L}) = \sum_{i=1}^{I} \alpha_i \gamma_1\left(\frac{ L_1}{\theta_1} + \sum_{j=2}^{i} \frac{L_j - L_{j-1}}{\theta_j}\right).
\end{equation}

\begin{algorithm}[!t]
\DontPrintSemicolon
\SetAlgoLined
\KwIn{Iteration rounds $Itr_{max}$, threshold for convergence $\varepsilon$, initial guess for $\mathbf{L}^0$ and $\lambda^0$, historical AIGC service quality data $\{\tilde\xi_n| n \in [N]\}$ and update step size of $\mathbf{L}$-block and $\lambda$-block, i.e., $\eta_L$ and $\eta_{\lambda}$}
\KwOut{Robust pricing contract bundles $(\mathbf{L}^*, \mathbf{R}^*)$ for AIGC tasks offloading}
\BlankLine
\tcc{Stage 1: Initial setup}
Acquire setup of the contract theory model, including contract type $\{\theta_i| i \in [I]\}$ and distribution $\{\alpha_i| i \in [I]\}$\;
Acquire setup of the DRO model, including $\epsilon$ via (\ref{eq:9}) and support set of $\Xi$, i.e., $\underline{\xi}$ and $\bar{\xi}$\;
Initialize hyperparameters of BCD algorithm, iteration rounds $t \leftarrow 0$, objective value ${\Omega}^* \leftarrow -\infty$\;
\tcc{Stage 2: BCD iteration}
\ForEach{$itr \in [Itr_{max}]$}{
    \tcc{Update $\mathbf{s}$-block}
    $\{\xi^*_n| n \in [N]\} \leftarrow$ Invoke Algorithm \ref{alg:3} \;
    \tcc{Update $\mathbf{L}$-block}
    Calculate the gradient of $\mathbf{L}$ as per (\ref{eq:47}) \;
    Update $\mathbf{L}$-block via (\ref{eq:48}) \;
    Utilize the ironing and bunching method to enforce the monotonicity of $\mathbf{L}$\;
    \tcc{Update $\mathbf{L}$-block}
    Calculate the gradient of $\lambda$ as per (\ref{eq:49}) \;
    Update $\lambda$-block via (\ref{eq:50}) \;

    $\Omega^t \leftarrow$ Invoke Algorithm \ref{alg:3} \;
    \If{$|\Omega^* - \Omega^t| \leq \varepsilon$}{
        break\;
    }\Else{
        Update objective value, $\Omega^* = \Omega^t$ \;
    }
}
Calculate $\mathbf{R}^*$ as per (\ref{eq:33}) \;
\Return{$(\mathbf{L}^*, \mathbf{R}^*)$}\;
\caption{Block coordinate descent-driven contract theory}
\label{alg:1}
\end{algorithm}

In light of we intend to maximize $-\lambda \epsilon + \frac{1}{N} \sum_{n=1}^{N} s_n$, each $s_n$ is binding for (\ref{P3_constraint_1}), i.e.,
\begin{equation} \label{eq:43}
    \underbrace{\left(\min_{\xi \in [\underline{\xi}, \bar{\xi}]} f_n(\xi;\mathbf{L}, \lambda) \right)}_{\colon \phi_n(\mathbf{L}, \lambda)} - g(\mathbf{L}) = s_n.
\end{equation}
Therefore, (\ref{eq:34}) can be further simplified as
\begin{subequations} \label{eq:44}
	\begin{align}
		\sup_{\mathbf{L}, \lambda} & \quad -\lambda \epsilon + \frac{1}{N} \sum_{n=1}^{N} \left(\phi_n(\mathbf{L}, \lambda)- g(\mathbf{L})\right)   \label{Problem_4} \\
		  s.t. & \quad L_1 \leq \dots \leq L_i \leq \dots \leq L_I, \label{P4_constraint_1} \\
          & \quad \lambda \geq 0. \label{P4_constraint_2}
	\end{align}
\end{subequations}

Considering the nested structure of (\ref{eq:44}), we intend to utilize the BCD Algorithm to resolve it. Specifically, we will categorize the variables in (\ref{eq:44}) into three blocks, which are $\mathbf{s}$-block, $\mathbf{L}$-block and $\lambda$-block.

Notably, the $\mathbf{s}$-block is fixed once $\mathbf{L}$ and $\lambda$ are fixed and thus the key of $\mathbf{s}$-block is $\phi_n(\mathbf{L}, \lambda)$. Given that $ln(\gamma_2\xi + \gamma_3L_i)$ increases in $\xi$ and $\lambda \| \xi - \hat{\xi}_n \|$ is a V-shaped piecewise linear function in $\xi$, the optimal point of $\phi_n(\mathbf{L}, \lambda)$ is achieved in one of the following cases $\{\underline{\xi}, \bar{\xi}, \tilde{\xi}_n, \xi^{p}_n\}$. Here, $\xi^p \in (\underline{\xi}, \tilde{\xi}_n)$ and defined as
\begin{equation} \label{eq:45}
    \left. \frac{d}{d\xi} \left( \sum_{i=1}^{I} \alpha_i \ln (\gamma_2 \xi + \gamma_3 L_i^{itr}) \right) \right|_{\xi = \xi^p_n} = \lambda.
\end{equation}

\begin{algorithm}[!t]
\DontPrintSemicolon
\SetAlgoLined
\KwIn{$\mathbf{L}$-block, $\lambda$-block and sampled AIGC service quality data $\tilde\xi_n$}
\KwOut{Optimal value and results of $\phi_n(\mathbf{L}, \lambda)$, i.e., $f_{{\xi}^*_n}$ and $\xi^*_n$}
\BlankLine
Initialize hyperparameters of the contract theory and DRO model\;
$f_{\bar{\xi}} \leftarrow$ Set $\xi$ as $\bar{\xi}$ and evaluate the value of (\ref{eq:41})\;
$f_{\underline{\xi}} \leftarrow$ Set $\xi$ as $\underline{\xi}$ and evaluate the value of (\ref{eq:41})\;
\If{ $\underline{\xi} \leq \tilde \xi_n \leq \bar{\xi}$}
{
    $f_{\tilde{\xi}_n} \leftarrow$ Set $\xi$ as $\tilde{\xi}_n$ and evaluate the value of (\ref{eq:41})\;
    $\xi^{p}_n \leftarrow$ Utilize the method of binary search to resolve (\ref{eq:46}) in $(\underline{\xi}, \tilde{\xi}_n)$\;
    $f_{{\xi}^p_n} \leftarrow$ Set $\xi$ as ${\xi}^p_n$ and evaluate the value of (\ref{eq:41})\;
}
$(f_{{\xi}^*_n}, \xi^*_n) \leftarrow$ Take the minimum value and the corresponding index over $\{f_{\bar{\xi}}, f_{\underline{\xi}}, f_{\tilde{\xi}_n}, f_{{\xi}^p_n}\}$\;

\Return{$(f_{{\xi}^*_n}, \xi^*_n)$}\;
\caption{$\xi^*$ solver}
\label{alg:2}
\end{algorithm}

With (\ref{eq:45}), we can deduce that the $\xi^p_n$ is obtained by solving
\begin{equation} \label{eq:46}
    \sum_{i=1}^{I} \frac{\alpha_i\gamma_2}{\gamma_2\xi + \gamma_3L_i^{itr}} = \lambda.
\end{equation}
It should be noted that $\xi^p_n$ is valid if and only if $\xi^p_n \in (\underline{\xi}, \tilde{\xi}_n)$.

Subsequently, with fixed $\mathbf{L}$ and $\lambda$ and taking the minimum of $\phi_n(\mathbf{L}, \lambda$ over $\{\underline{\xi}, \bar{\xi}, \tilde{\xi}_n, \xi^{p}_n\}$ and obtaining $\xi^*_n$, we can obtain the updated $\mathbf{s}$-block.

Regarding the $\mathbf{L}$-block update, we will fix $\lambda$ and derive the gradient ascent formula for $\mathbf{L}$, i.e., the gradient of (\ref{Problem_4}) with respect to $\mathbf{L}$, which is defined as
\begin{equation} \label{eq:47}
    \nabla L_i = \frac{1}{N}\sum_{n=1}^{N} \left(\frac{\alpha_i\gamma_3}{\gamma_2\xi^*_n + \gamma_3L_i} - \frac{\alpha_i \gamma_1}{\theta_i} \right).
\end{equation}
Here, for simplicity, we overlook the dependence between $\xi^*_n$ and $\mathbf{L}$, and therefore (\ref{eq:47}) is an approximate gradient of $\mathbf{L}$. The update formula for $\mathbf{L}$-block is defined as
\begin{equation} \label{eq:48}
    L_i^{itrt+1} = L_i^{itr} + \eta_L \nabla L_i,
\end{equation}
where $\eta_L$ is the updating step of $L_i$ and $itr$ and $itr+1$ indicates the updating round of the BCD algorithm. Notably, the monotonicity constraint should be satisfied when update $\mathbf{L}$, thereby the bunch and ironing method \cite{gao2011spectrum} is implemented to revise $\mathbf{L}$ after each round of $\mathbf{L}$-block update.

Analogously, regarding the $\lambda$-block update, we will fix $\mathbf{L}$ and derive the gradient ascent formula for $\lambda$, which is defined as
\begin{equation} \label{eq:49}
     \nabla \lambda = -\epsilon + \frac{1}{N}\sum_{n=1}^N \|\xi_n^* - \tilde\xi_n\|.
\end{equation}
Accordingly, the $\lambda$-block update formula is defined as
\begin{equation} \label{eq:50}
    \lambda^{itr+1} = \max(\lambda^{itr} + \eta_{\lambda} \nabla \lambda, ~ 0),
\end{equation}
where $\eta_{\lambda}$ is the updating step of $\lambda$.

\subsection{Algorithm Complexity Analysis} \label{sec:5.2}
For the sake of clarity, we present the BCD-Driven Contract Theory in Algorithm \ref{alg:1}, which includes two subroutines, Algorithms \ref{alg:2} and \ref{alg:3}.

The complexity of Algorithm \ref{alg:1} is $\mathcal{O}\left(Itr_{max}\left(\frac{I(I-1)}{2} + N(B+I)\right)\right)$, where $B$ is the number of iteration rounds of the binary search used to solve (\ref{eq:46}). Specifically, the total number of iterations of the BCD algorithm contributes to the factor $Itr_{max}$. The invocation of Algorithm \ref{alg:3} incurs a complexity of $\left(\frac{I(I-1)}{2} + N(B+I)\right)$, where the calculation of $g(\cdot)$ accounts for $\frac{I(I-1)}{2}$ and the invocation of Algorithm \ref{alg:2} incurs a complexity of $N(B+I)$. Additionally, the complexity of updating the $\mathbf{L}$-block and $\lambda$-block is $NI$, which is less than $\left(\frac{I(I-1)}{2} + N(B+I)\right)$. Therefore, the total complexity of Algorithm \ref{alg:1} is $\mathcal{O}\left(Itr_{max}\left(\frac{I(I-1)}{2} + N(B+I)\right)\right)$.

\begin{algorithm}[!t]
\DontPrintSemicolon
\SetAlgoLined
\KwIn{$\mathbf{L}$-block, $\lambda$-block and historical AIGC service quality data $\{\tilde\xi_n| n \in [N]\}$}
\KwOut{Objective value of (\ref{eq:44}) $\Omega$ and optimal value of $\xi$, $\{\xi^*_n| n \in [N]\}$}
\BlankLine
Initialize parameters, including $\{\xi^*_n = 0 | n \in [N]\}$ and $\bar{s} = 0$ \;
Calculate $g(\cdot)$ as per (\ref{eq:42})\;
\ForEach{$n \in [N]$}{
    $(f_{{\xi}^*_n}, \xi^*_n) \leftarrow$ Invoke Algorithm \ref{alg:2}\;
    Utilize $g(\cdot)$ to calculate $s_n$ as per (\ref{eq:43})\;
    $\xi^*_n \leftarrow \xi^*_n$ \;
    $\bar{s} += \frac{s_n}{N}$ \;
}
Calculate the objective value, $\Omega = -\lambda \epsilon + \bar{s}$ \;
\Return{$\Omega$, $\{\xi^*_n| n \in [N]\}$}\;
\caption{Objective value calculate}
\label{alg:3}
\end{algorithm}

\begin{figure*}[!t]
    \centering
    \begin{minipage}[t]{0.32\textwidth}
        \centering
        \includegraphics[width=\linewidth]{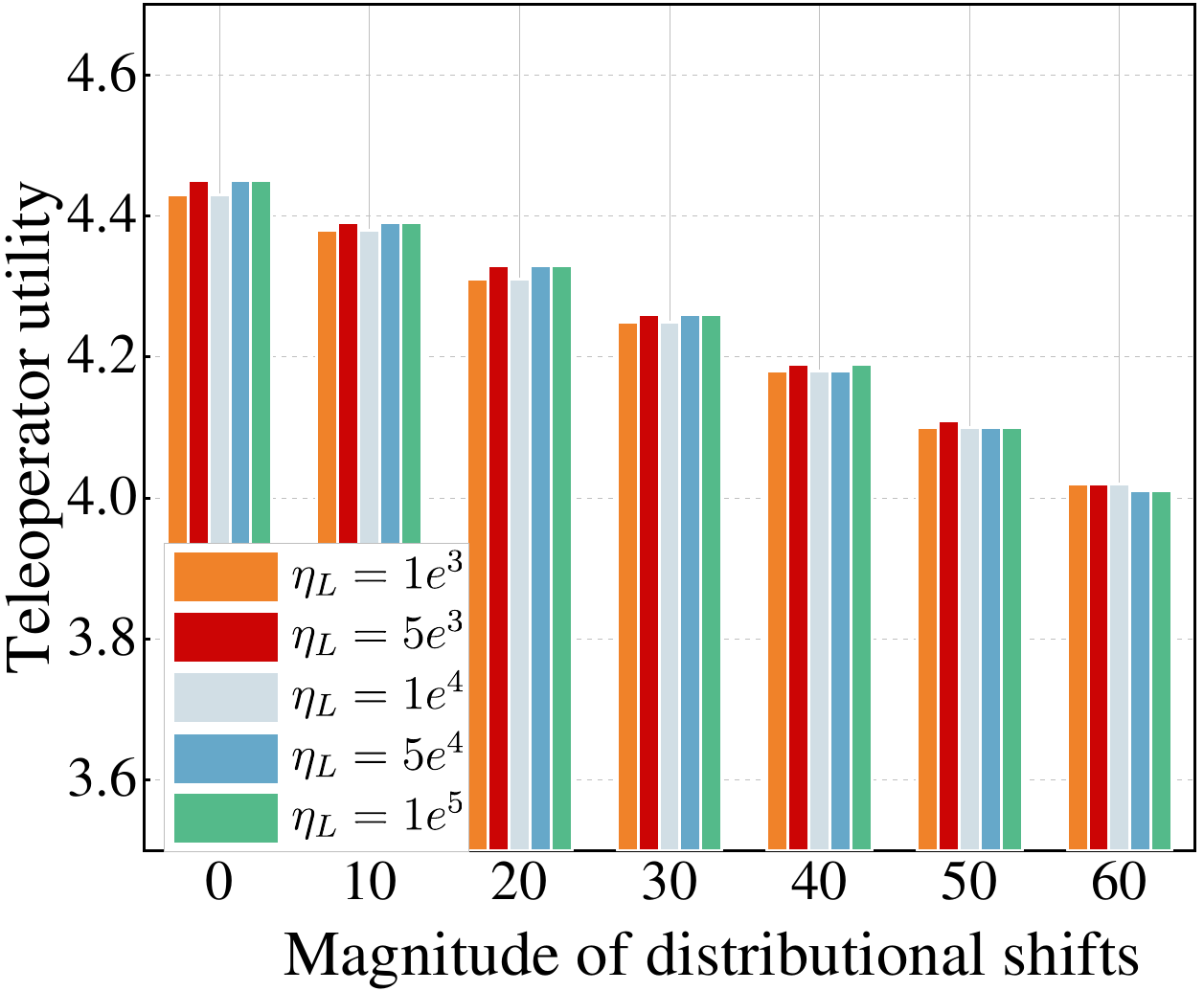}
        \caption{Teleoperator utility versus the magnitude of evaluation data distribution shifts under varying latency updating steps $\eta_L$.}
        \label{fig3}
    \end{minipage}
    \hfill
    \begin{minipage}[t]{0.32\textwidth}
        \centering
        \includegraphics[width=\linewidth]{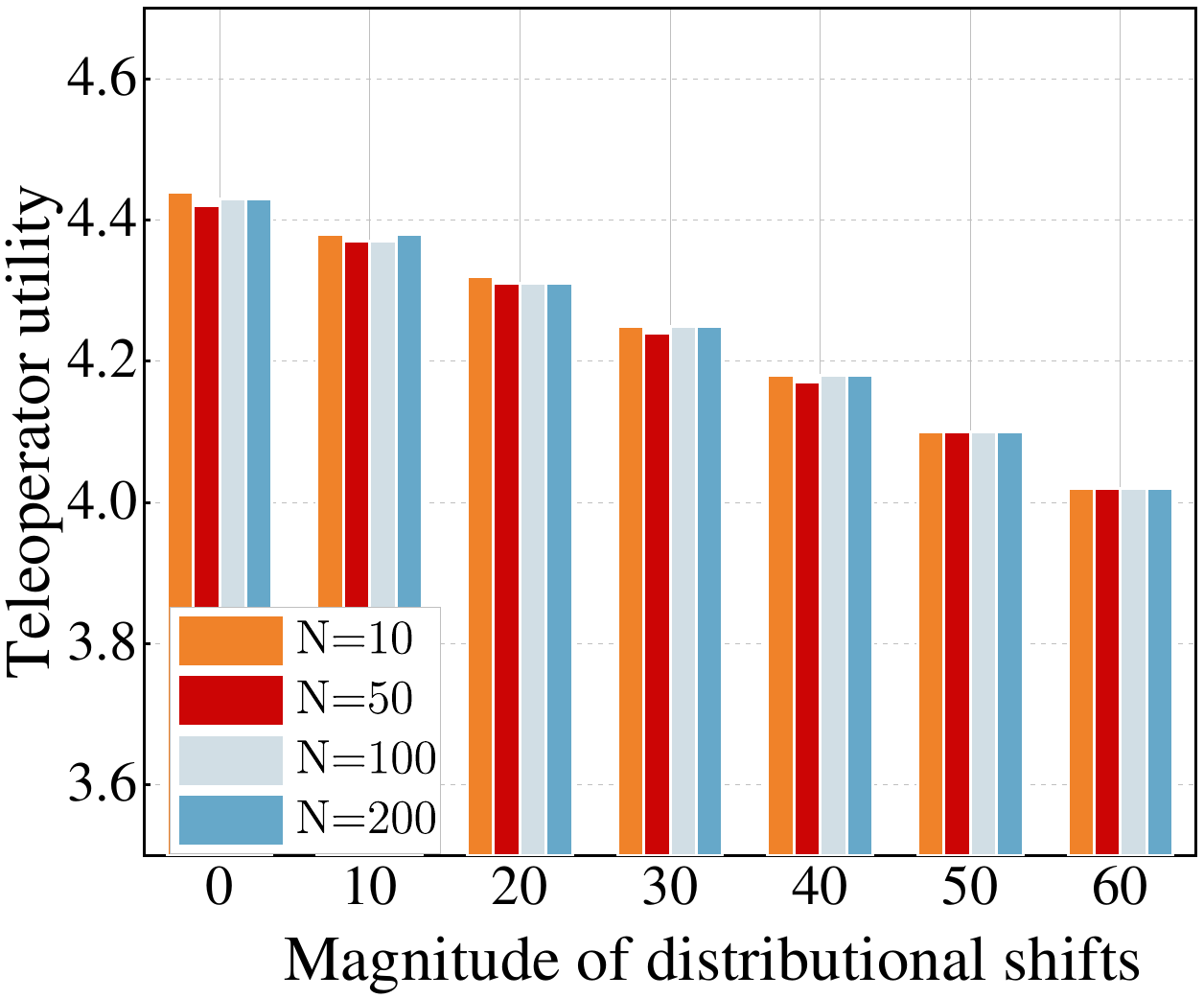}
        \caption{Teleoperator utility versus the magnitude of evaluation data distribution shifts when the number of training data $N$ is varied.}
        \label{fig4}
    \end{minipage}
    \hfill
    \begin{minipage}[t]{0.32\textwidth}
        \centering
        \includegraphics[width=\linewidth]{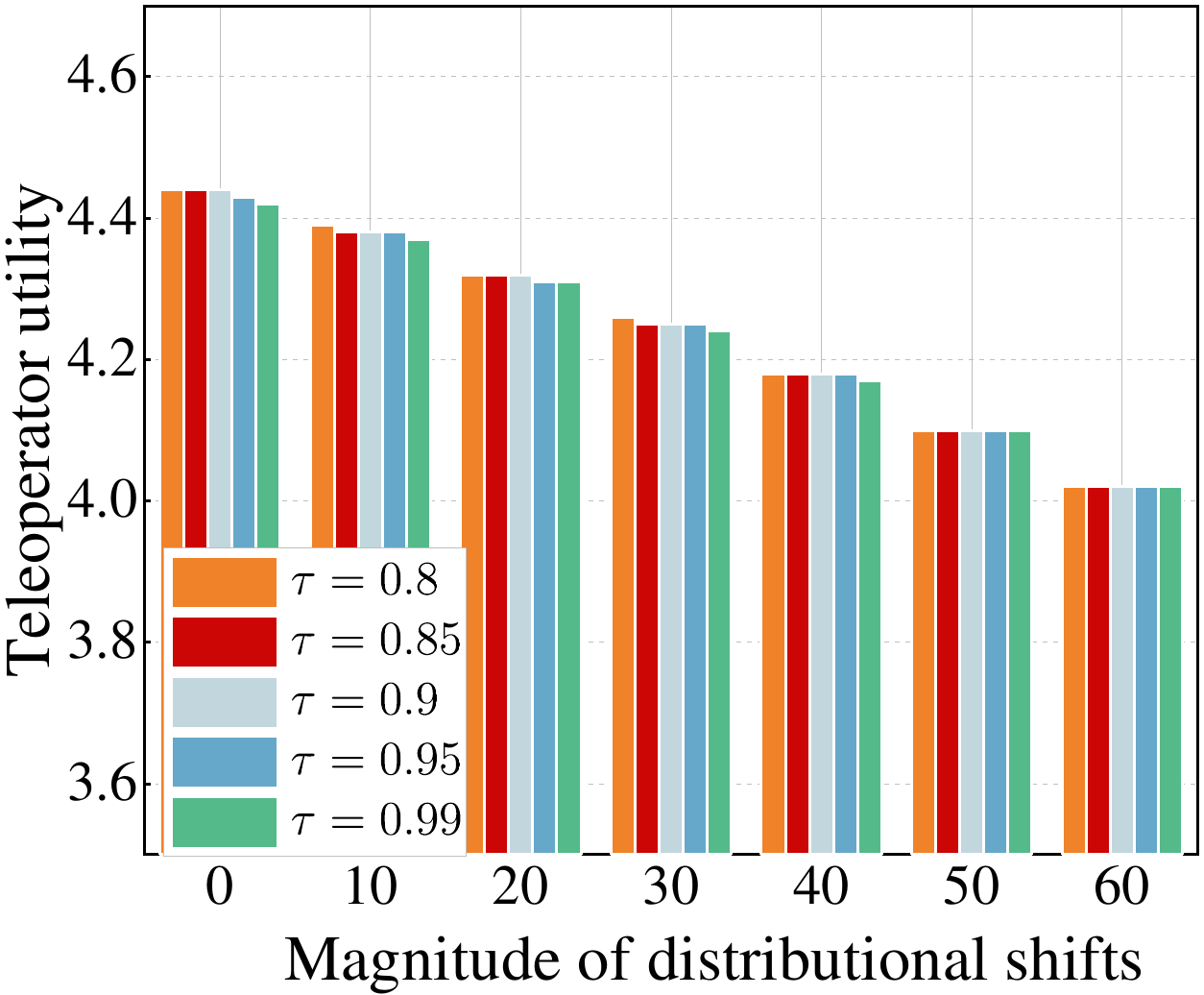}
        \caption{Teleoperator utility versus the magnitude of evaluation data distribution shifts under varying confidence levels $\tau$.}
        \label{fig5}
    \end{minipage}
\end{figure*}

\begin{figure*}[!t]
    \centering
    \begin{minipage}[t]{0.32\textwidth}
        \centering
        \includegraphics[width=\linewidth]{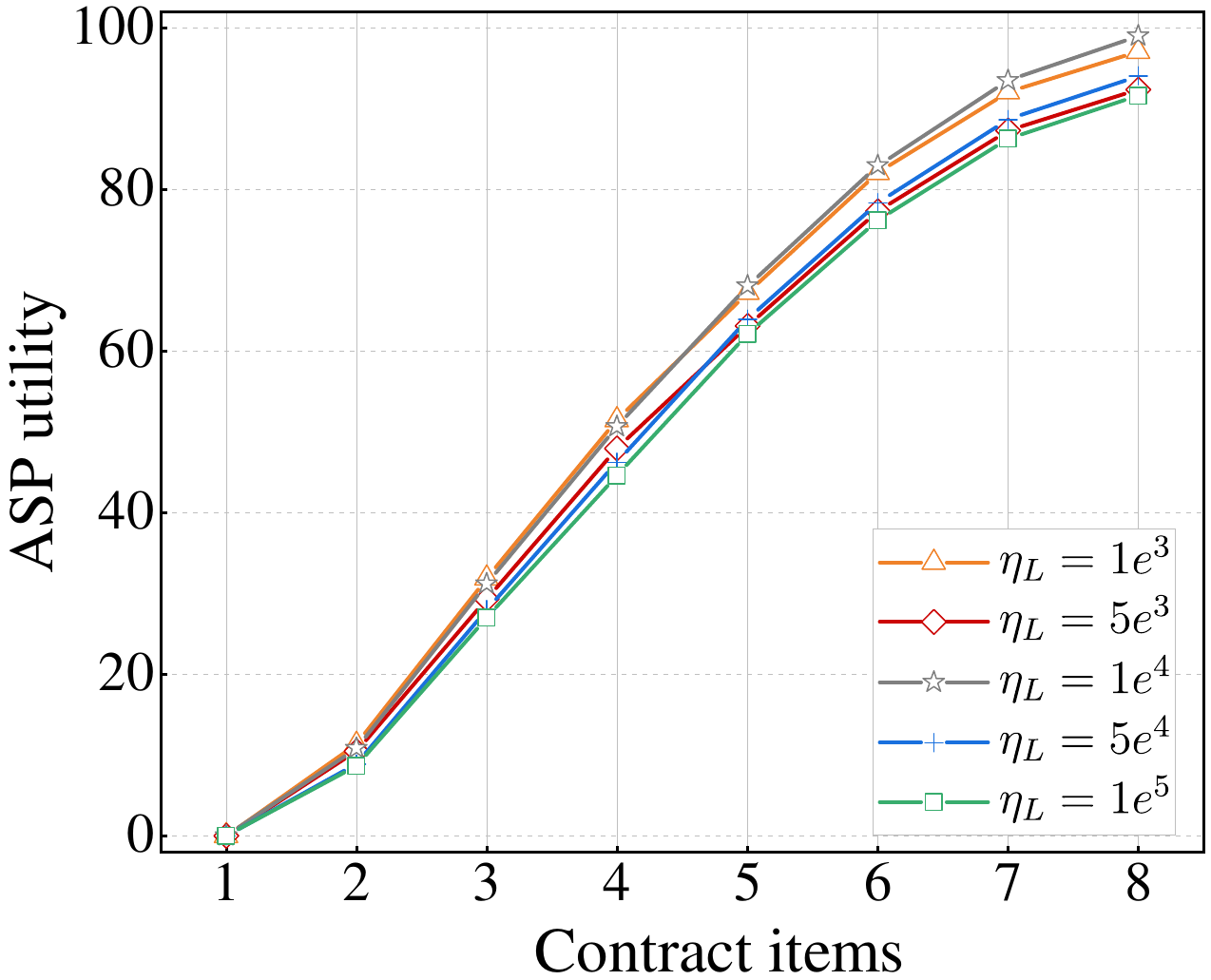}
        \caption{ASP utility versus the type of contract items under varying latency updating steps $\eta_L$.}
        \label{fig6}
    \end{minipage}
    \hfill
    \begin{minipage}[t]{0.32\textwidth}
        \centering
        \includegraphics[width=\linewidth]{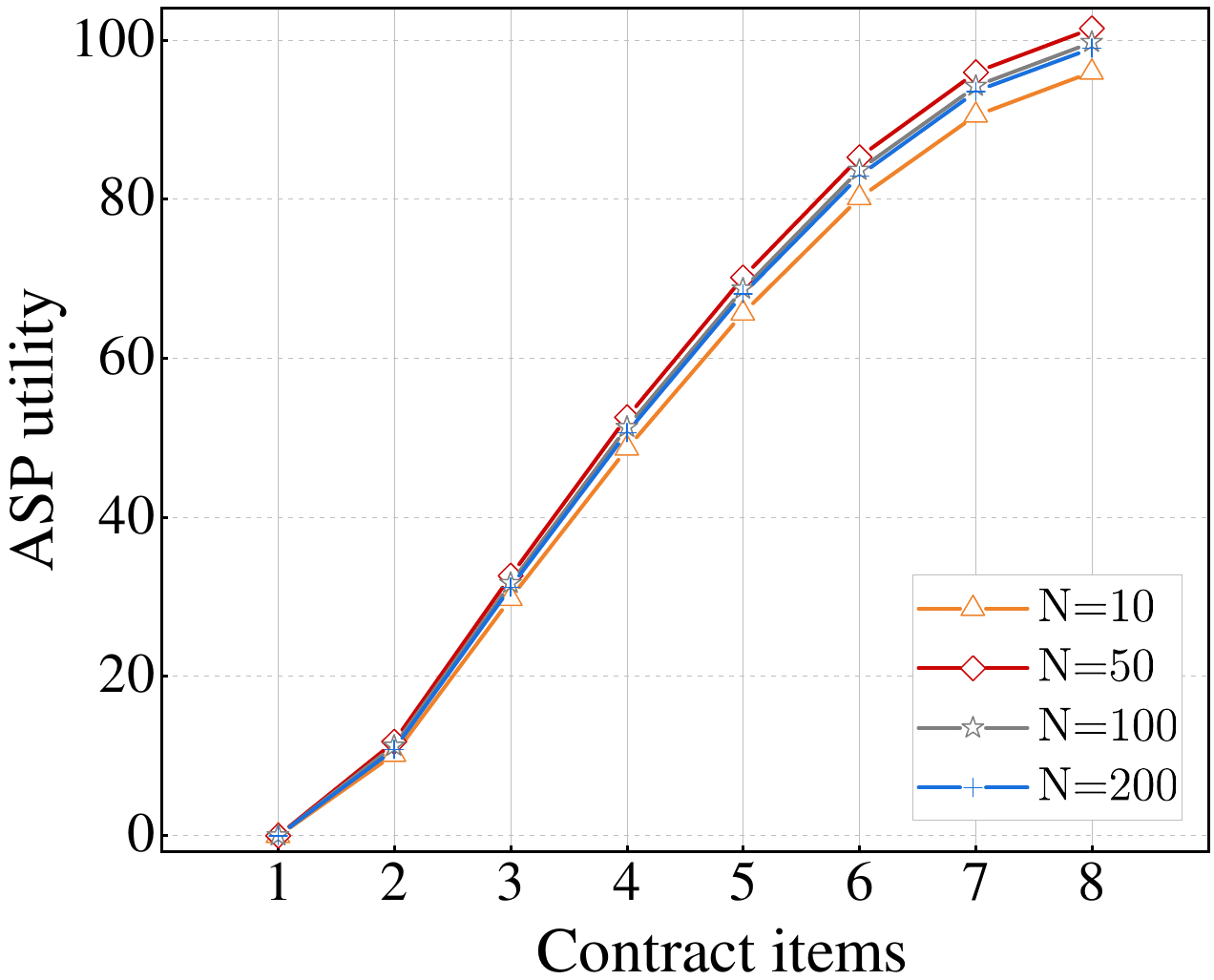}
        \caption{ASP utility versus the type of contract items when the number of training data $N$ is varied.}
        \label{fig7}
    \end{minipage}
    \hfill
    \begin{minipage}[t]{0.32\textwidth}
        \centering
        \includegraphics[width=\linewidth]{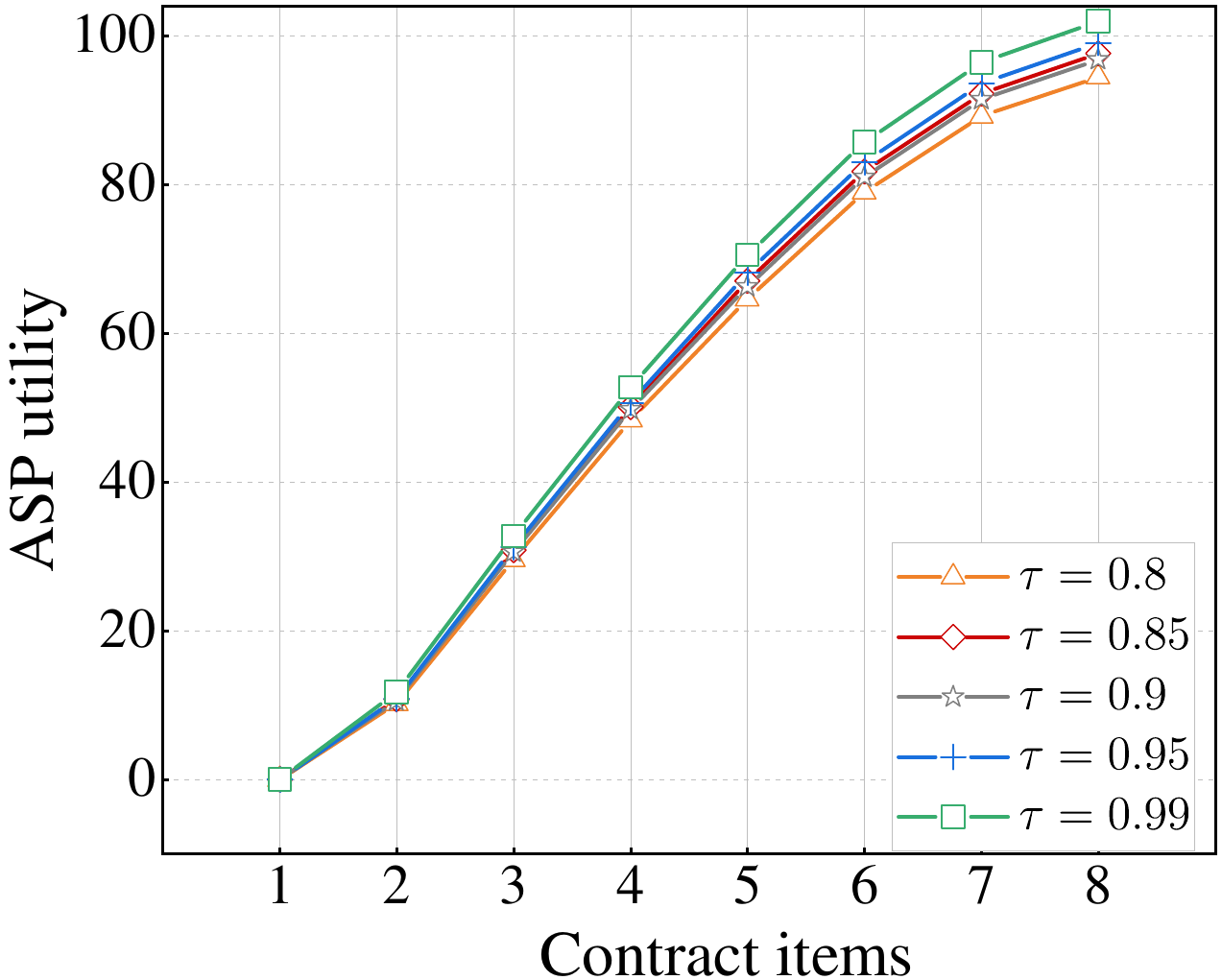}
        \caption{ASP utility versus the type of contract items under varying confidence levels $\tau$.}
        \label{fig8}
    \end{minipage}
\end{figure*}

\section{Experimental Evaluation} \label{sec:6}
In this section, we first introduce the experimental configurations, including the data, parameter settings, benchmarks, and evaluation metrics, in Section \ref{sec:6.1}. Next, in Section \ref{sec:6.2}, we demonstrate the effectiveness of our proposed method by varying key hyperparameters. Subsequently, in Section \ref{sec:6.3}, we validate the robustness of our proposed method by comparing it with baselines in terms of ASP utility and the robustness of teleoperator utility. Lastly, in Section \ref{sec:6.4}, we assess the convergence of our proposed method through experimental results.

\subsection{Configurations} \label{sec:6.1}
The GDM-based AIGC model adopted in this paper is trained on 526 paired low-light and normal-light images from our developed unity project. Additionally, we collect another 250 paired images to validate the performance of the AIGC model. Notably, the performance score of the AIGC results is based on a combination of LPIPS \cite{zhang2018unreasonable} and SSIM \cite{wang2004image}. We then split the performance results into a 200:50 ratio, where 200 results are used as training data for the DRO-based contract theory, and the remaining 50 results are used as evaluation data.

The parameters used in this paper are categorized into three parts: the contract theory model, the DRO model, and the BCD algorithm. First, for the contract theory model, we consider a general case where the number of contract bundles is set to $I = 8$, and the corresponding willingness values $\{\theta_i | i \in [I]\}$ are set as $\{110, 140, 175, 200, 220, 235, 245, 250\}$, following settings similar to those in \cite{liu2024deep, wen2024diffusion}. Moreover, the probability of each type of ASP $\{\alpha_i | i \in [I]\}$ is randomly generated according to a Dirichlet distribution. For clarity and without loss of generality, the weighting coefficients are set as $\gamma_1 = 1$, $\gamma_2 = 1$, and $\gamma_3 = 1$. Second, regarding the DRO model, based on the evaluation results of the AIGC model, we set $\Xi = 40$, with $\bar{\xi} = 100$ and $\underline{\xi} = 60$. In addition, unless otherwise specified, we set the hyperparameters $\tau = 0.99$ and $N = 200$. Finally, for the BCD algorithm, we set the number of iteration rounds to $Itr_{max} = 1500$, the convergence threshold to $\varepsilon = 1e^{-3}$, the initial guess to $\mathbf{L}^0 = \{0 | i \in [I]\}$ and $\lambda^0 = 6$, and the updating step for $\lambda$ to $\eta_{\lambda} = 1e^{-3}$. By default, the key hyperparameter $\eta_{L}$ is set to $1e^4$.

\begin{figure*}[!t]
    \centering
    \begin{minipage}[t]{0.32\textwidth}
        \centering
        \includegraphics[width=\linewidth]{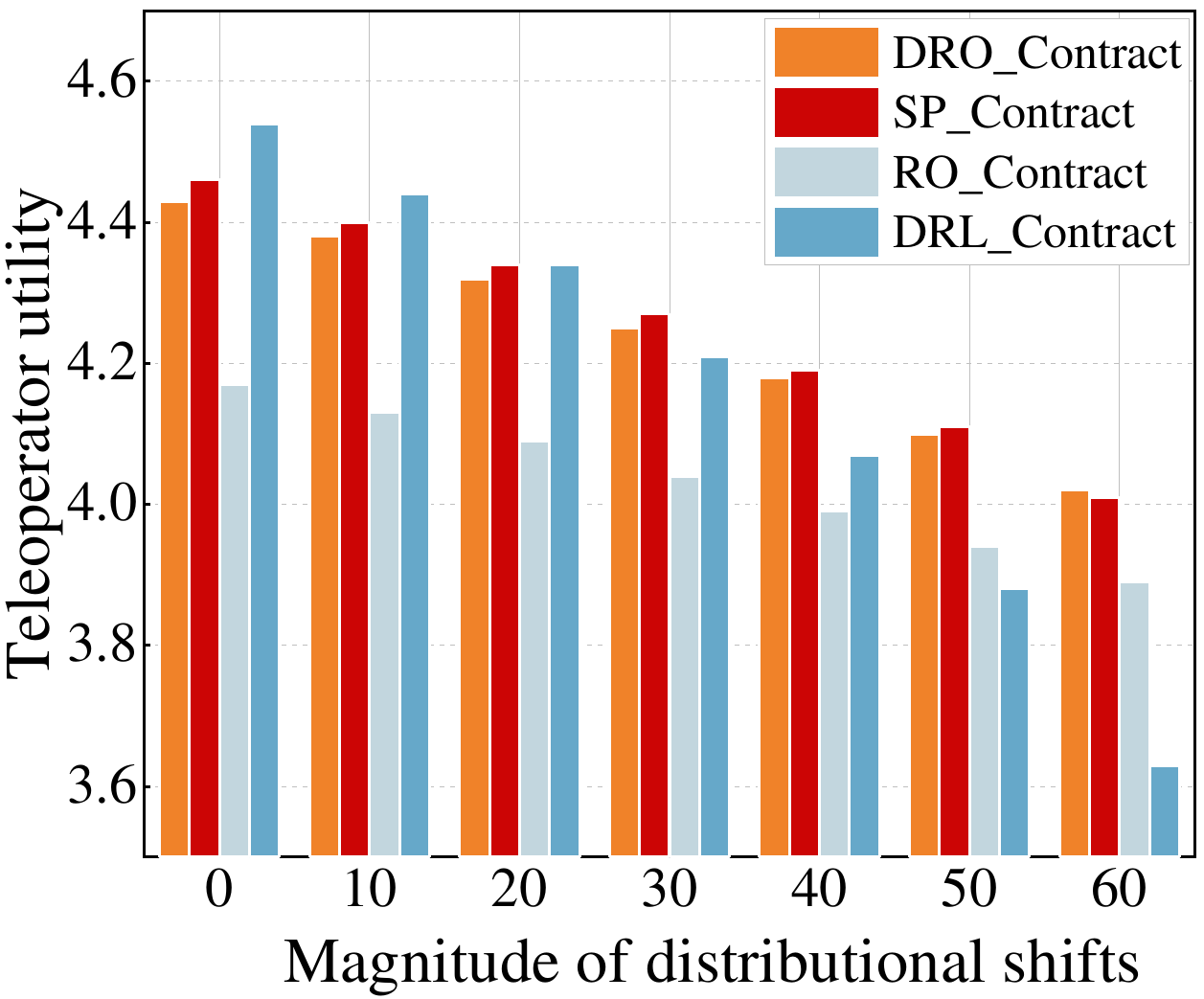}
        \caption{Comparison of teleoperator utility under different magnitudes of evaluation data distribution shifts with 0 extreme points.}
        \label{fig9}
    \end{minipage}
    \hfill
    \begin{minipage}[t]{0.32\textwidth}
        \centering
        \includegraphics[width=\linewidth]{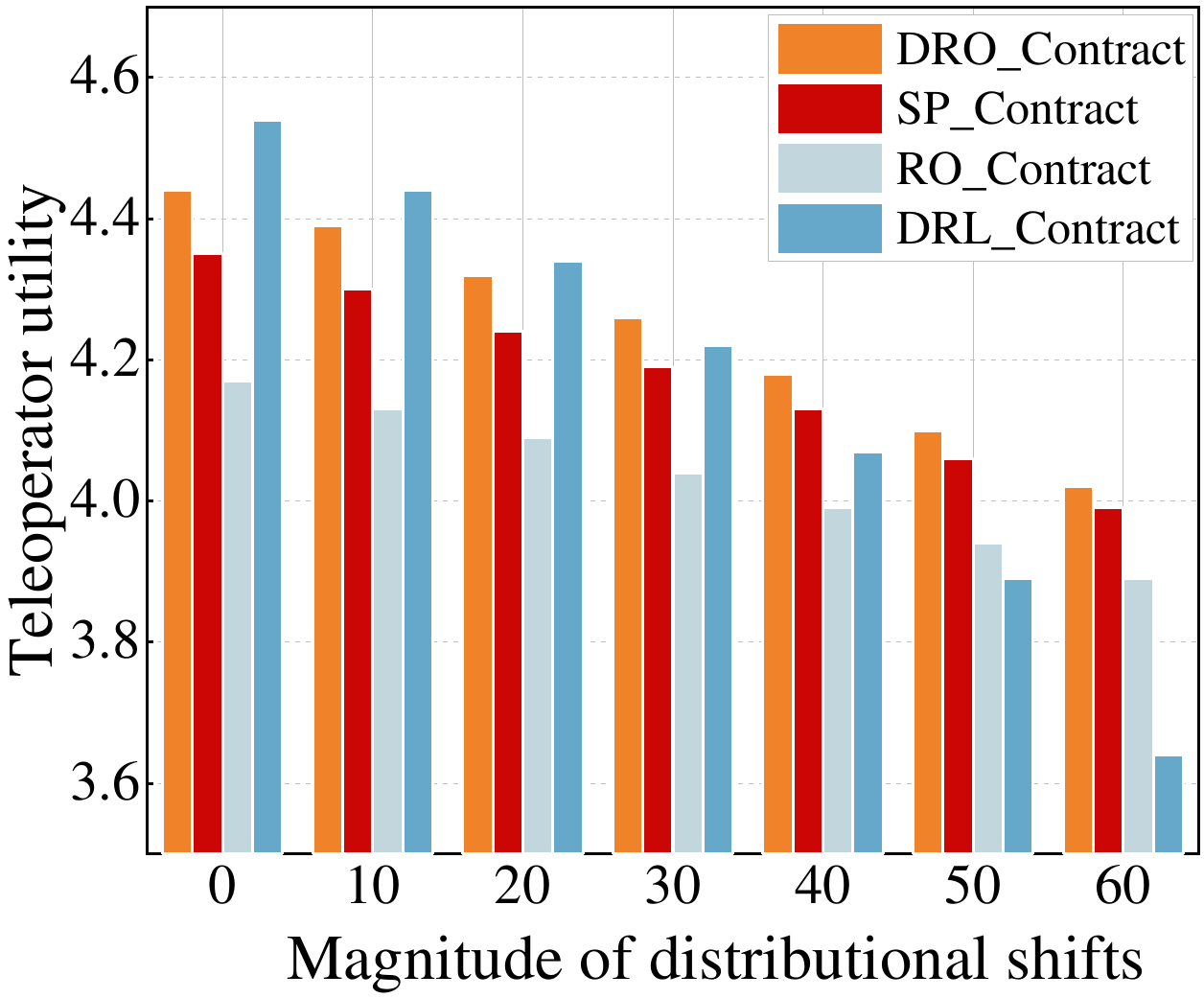}
        \caption{Comparison of teleoperator utility under different magnitudes of evaluation data distribution shifts with 50 extreme points.}
        \label{fig10}
    \end{minipage}
    \hfill
    \begin{minipage}[t]{0.32\textwidth}
        \centering
        \includegraphics[width=\linewidth]{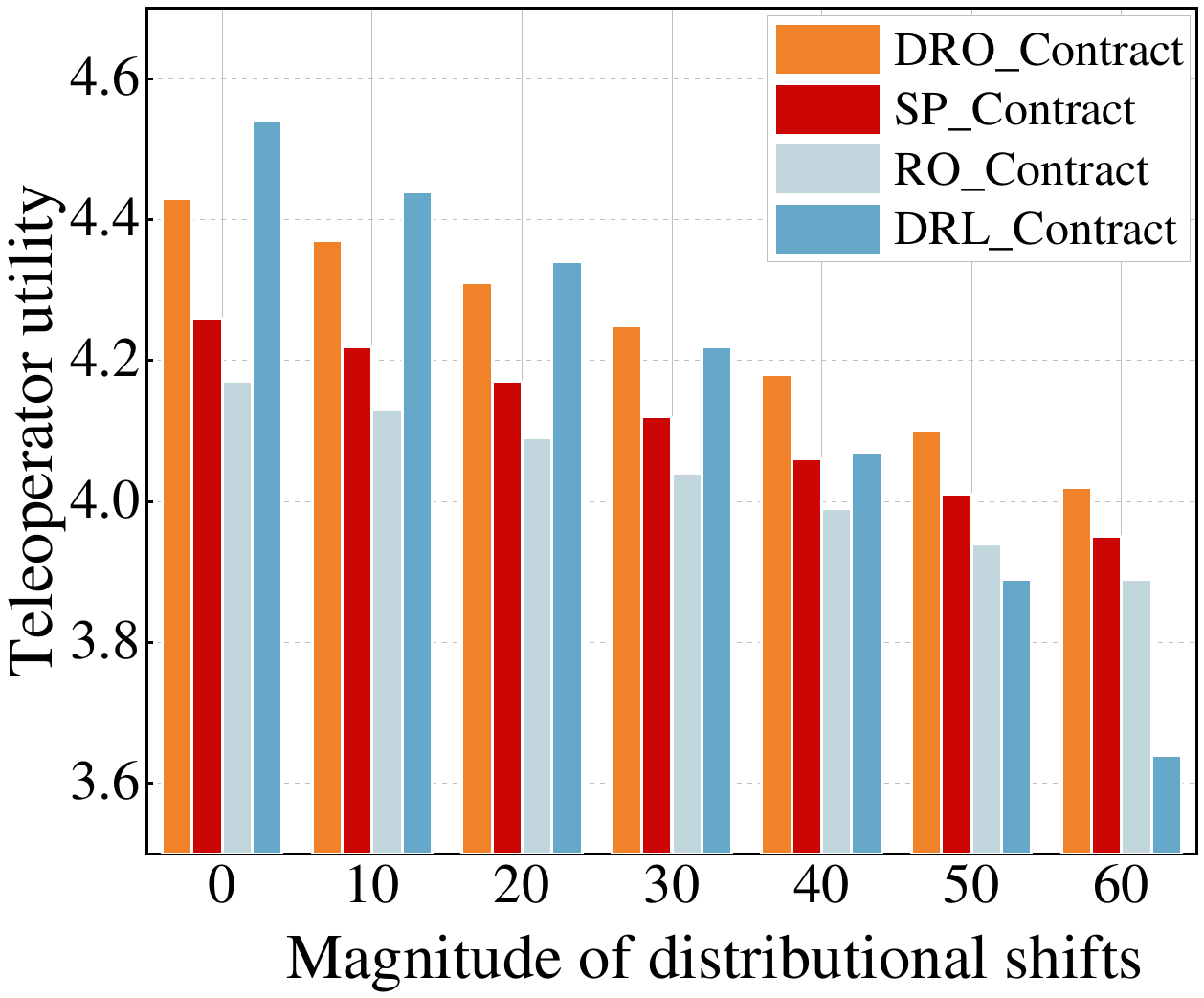}
        \caption{Comparison of teleoperator utility under different magnitudes of evaluation data distribution shifts with 100 extreme points.}
        \label{fig11}
    \end{minipage}
\end{figure*}

\begin{figure*}[!t]
    \centering
    \begin{minipage}[t]{0.32\textwidth}
        \centering
        \includegraphics[width=\linewidth]{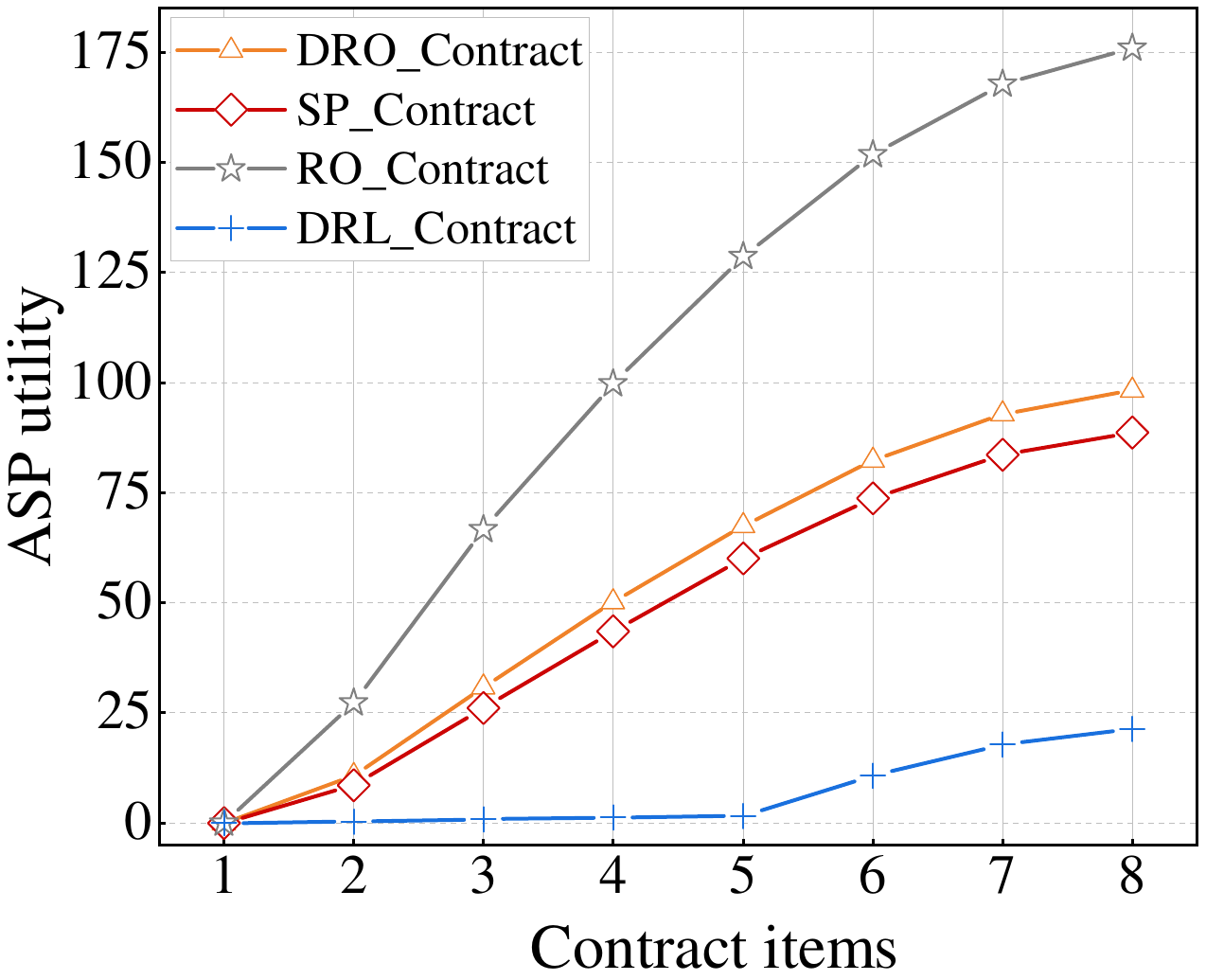}
        \caption{Comparison of ASP utility under different contract item types with 0 extreme points.}
        \label{fig12}
    \end{minipage}
    \hfill
    \begin{minipage}[t]{0.32\textwidth}
        \centering
        \includegraphics[width=\linewidth]{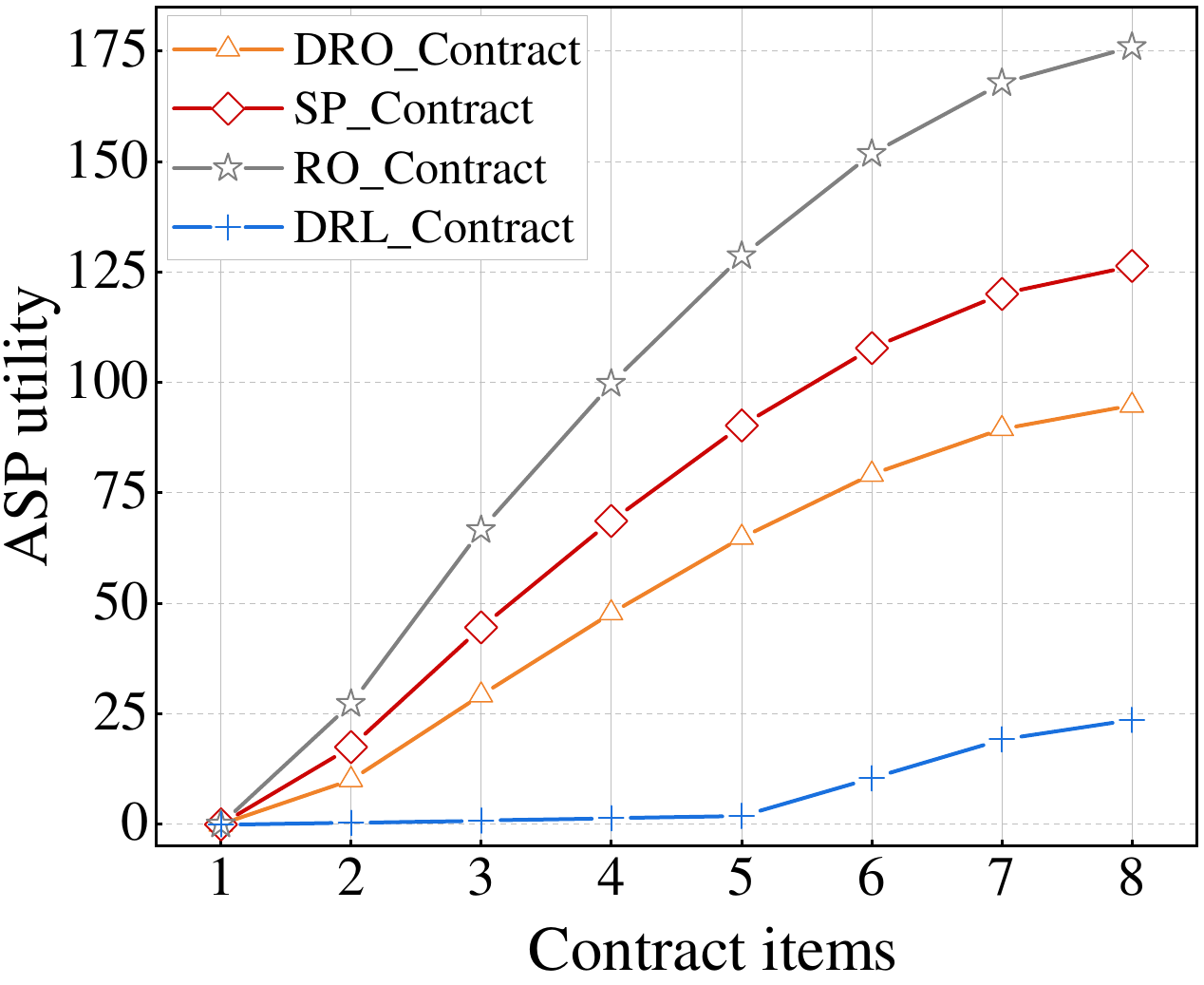}
        \caption{Comparison of ASP utility under different contract item types with 50 extreme points.}
        \label{fig13}
    \end{minipage}
    \hfill
    \begin{minipage}[t]{0.32\textwidth}
        \centering
        \includegraphics[width=\linewidth]{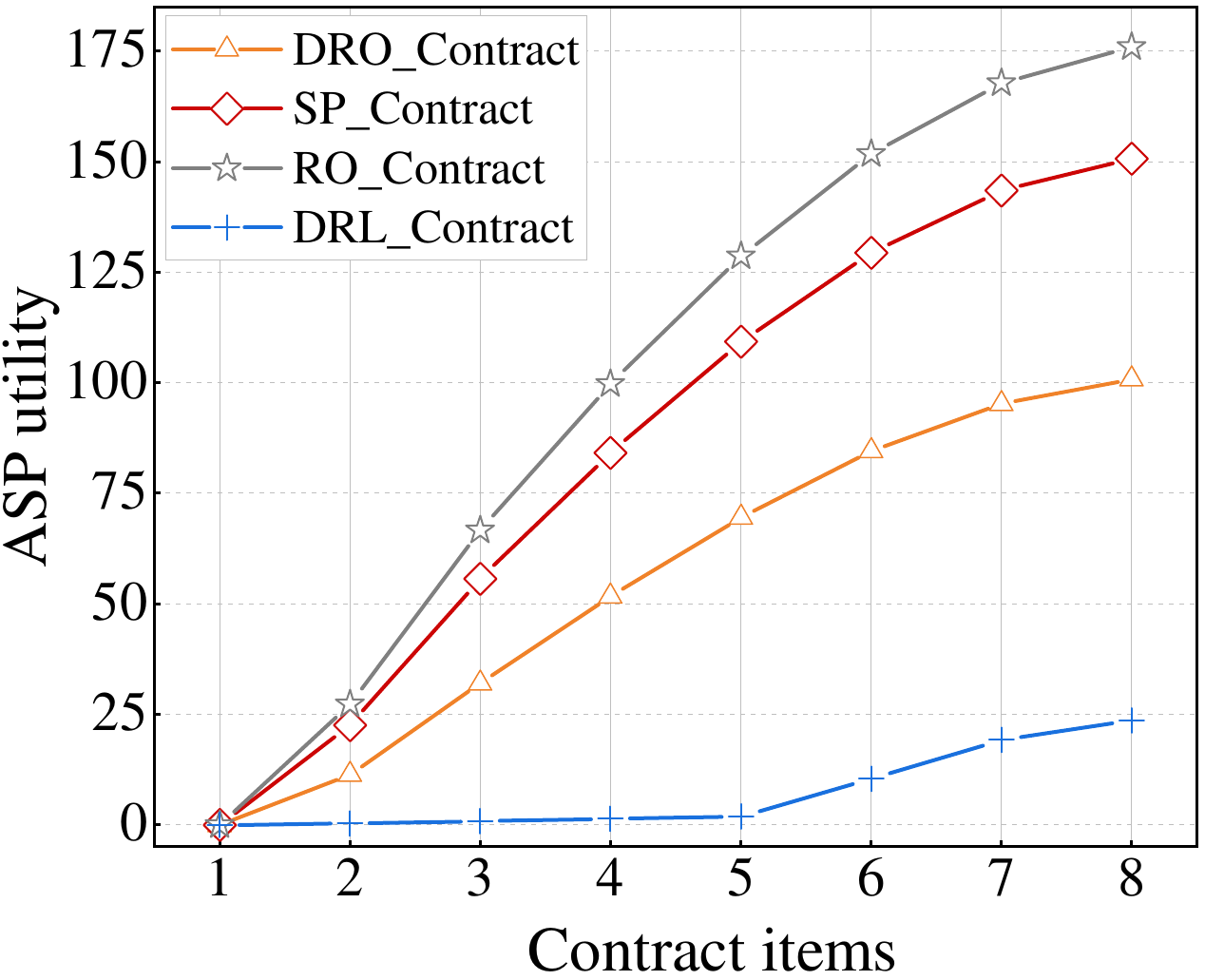}
        \caption{Comparison of ASP utility under different contract item types with 100 extreme points.}
        \label{fig14}
    \end{minipage}
\end{figure*}

To assess the effectiveness and robustness of our proposed DRO-based contract theory, we compare it with the following benchmarks:
\begin{enumerate}
    \item [1)] \textbf{SP\_Contract} \cite{schneider2006stochastic}: This baseline adopts stochastic programming to handle the uncertainty in the contract theory.

    \item [2)] \textbf{RO\_Contract} \cite{bertsimas2011theory}: This method applies robust optimization to address the uncertainty in the contract theory.
    
    \item [3)] \textbf{DRL\_Contract}\cite{wen2024learning, zhan2023, lotfi2024semantic}: This baseline leverages DRL to obtain converged contract bundles, using the Proximal Policy Optimization (PPO) algorithm.
\end{enumerate}
Our proposed method is designated as DRO\_Contract, thereby in line with baselines. Notably, all of the baselines will utilize the collected 200 data samples to obtain the finalized contract bundles and all simulations are conducted via Python and PyTorch.

To assess the performance of our proposed method in addressing Q1, the metrics used in this paper are the utility of the teleoperator and ASP, in which the priority of the teleoperator utility is higher than the ASP utility. Furthermore, we also concentrate on the robustness of teleoperator utility under varying AIGC service quality. 

\subsection{Effectiveness of DRO-based Contract Theory} \label{sec:6.2}
In this section, we assess the effectiveness of our proposed method via identifying optimal key hyperparameters setting, including contract bundle updating step $\eta_L$, number of historical data samples $N$, and confidence level $\tau$. Specifically, we identify the optimal $\eta_L$ in the range of $\{1e^3, 5e^3, 1e^4, 5e^4, 1e^5\}$, the optimal $N$ in the range of $\{10, 50, 100, 200\}$, and the optimal $\tau$ in the range of $\{0.8, 0.85, 0.9, 0.95, 0.99\}$ in terms of the ASP utility and the robustness of the teleoperator utility.

Observing Figs. \ref{fig3} and \ref{fig6}, we found that the teleoperator utility is stable under the varying setting of $\eta_L$ and the optimal ASP utility is achieved when we set $\eta_L = 1e^4$. Therefore, we set $\eta_L = 1e^4$ by default in the rest of experiments. Moreover, it is worth noting that the ASP utility presented in Fig. \ref{fig6} satisfies the monotonicity constraint (\ref{P2_constraint_2}), thereby demonstrating the feasibility of derived contract bundles. Next, the results in Figs. \ref{fig4} and \ref{fig7} reveal that the teleoperator and ASP utility are both stable under varying settings of $N$, which demonstrates the robustness of our proposed method. Notably, when we set $N = 10$, the ASP utility degrades 5.8\% due to overly ambiguous distribution approximation with limited historical data. Similarly, the monotonicity is also held in Fig. \ref{fig7}. Here, without loss generality, we set $N = 200$ by default in the remaining experiments. Lastly, observing the results depicted in Figs. \ref{fig5} and \ref{fig8}, we found that varying $\tau$ will not affect the teleoperator utility but the ASP utility. Concretely, the ASP utility is increased along with the confidence level, the basic rationale is that more potential distributions will taken into consideration when optimizing the contract bundles when a higher confidence level is selected. Therefore, we set $\tau = 0.99$ by default in the rest of the experiments. Moreover, the results depicted in Fig. \ref{fig8} satisfy the monotonicity as well.

\subsection{Robustness of DRO-based Contract Theory} \label{sec:6.3}

\begin{figure*}[!t]
    \centering
    \begin{minipage}[t]{0.32\textwidth}
        \centering
        \includegraphics[width=\linewidth]{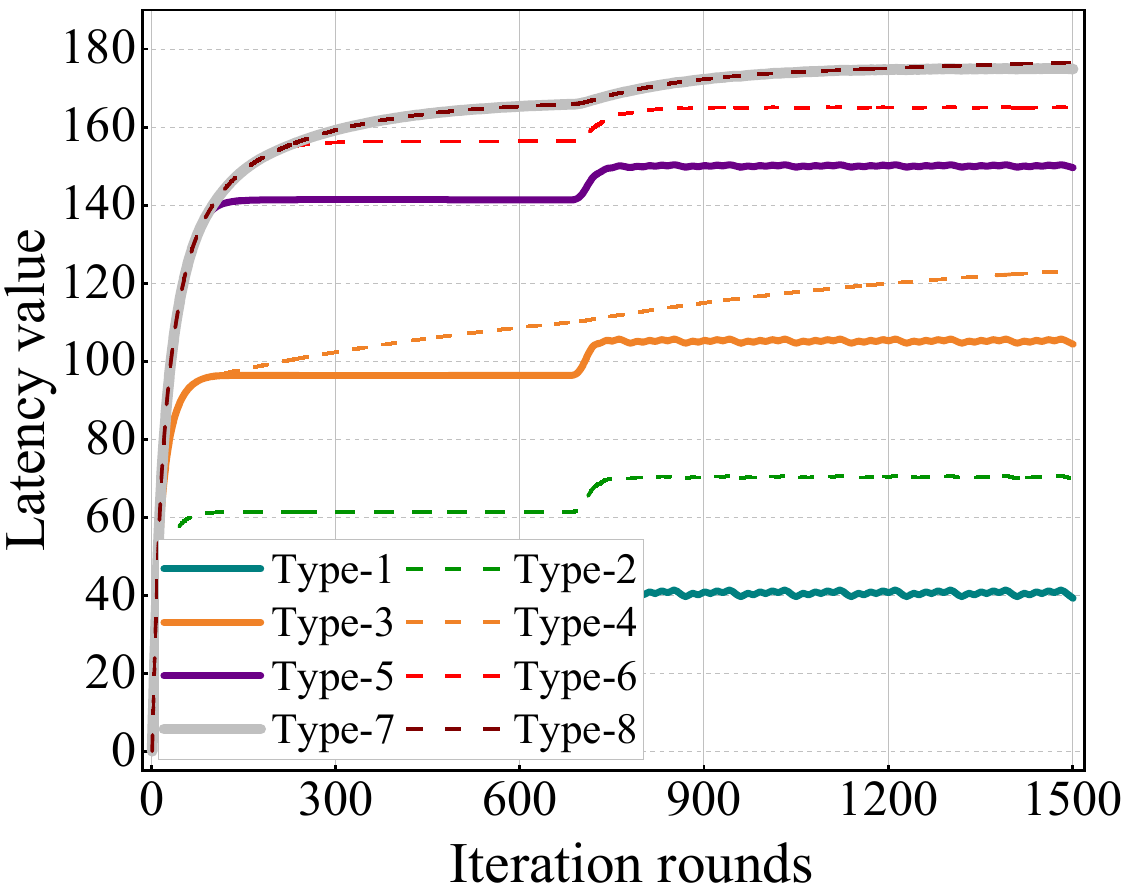}
        \caption{Convergence curve of contract bundles with 0 extreme points.}
        \label{fig15}
    \end{minipage}
    \hfill
    \begin{minipage}[t]{0.32\textwidth}
        \centering
        \includegraphics[width=\linewidth]{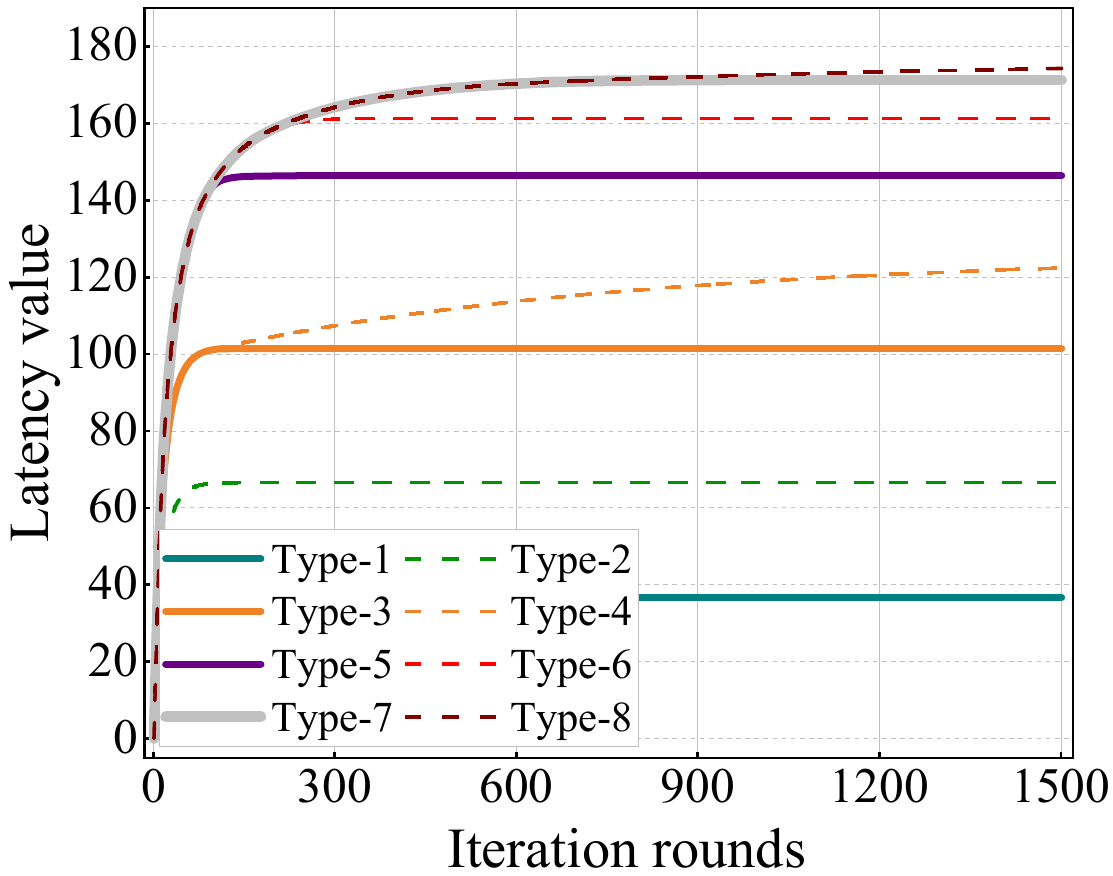}
        \caption{Convergence curve of contract bundles with 50 extreme points.}
        \label{fig16}
    \end{minipage}
    \hfill
    \begin{minipage}[t]{0.32\textwidth}
        \centering
        \includegraphics[width=\linewidth]{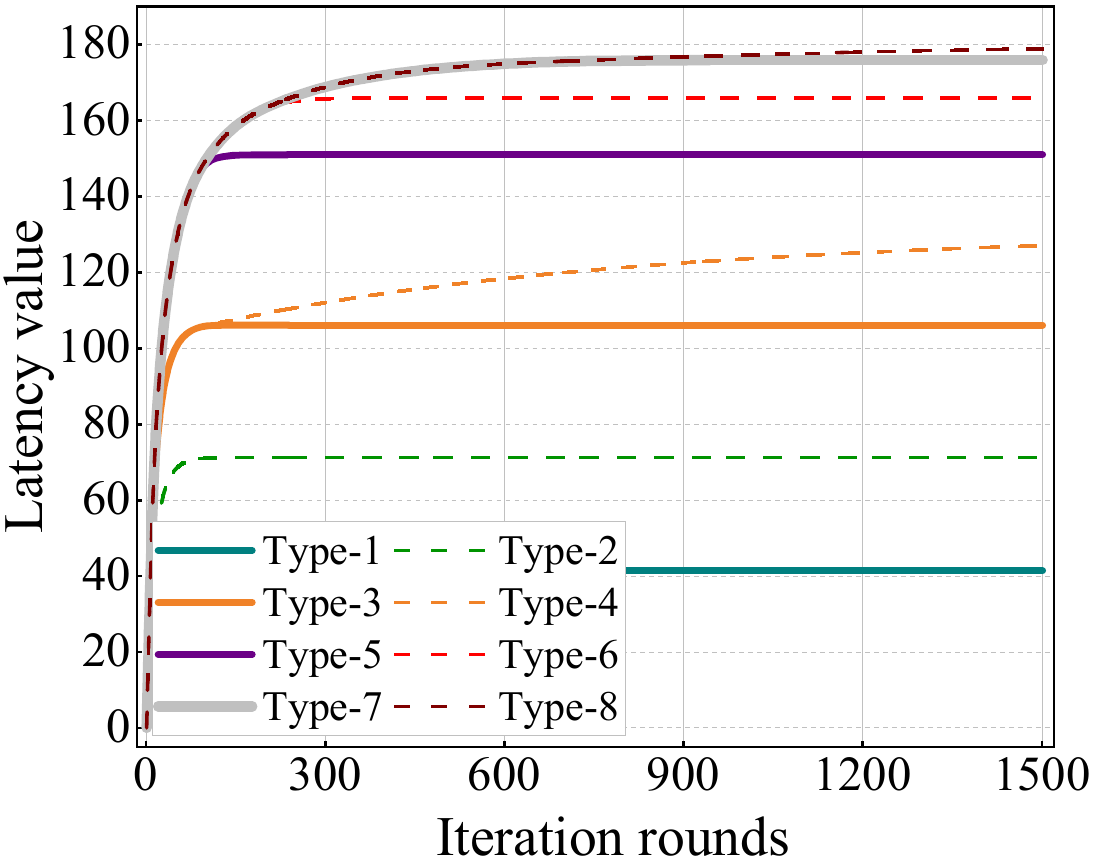}
        \caption{Convergence curve of contract bundles with 100 extreme points.}
        \label{fig17}
    \end{minipage}
\end{figure*}

In this section, we validate the robustness of our proposed method by comparing it with three baselines from two perspectives. First, we assess its robustness against varying magnitudes of evaluation data distribution shift. Second, we evaluate its robustness under different levels of training data quality.

Firstly, we assume the evaluation data distribution shift magnitude varies within $\{0, 10, 20, 30, 40, 50, 60\}$. As shown in Fig. \ref{fig9}, DRO\_Contract and SP\_Contract exhibit greater robustness compared to DRL\_Contract, while RO\_Contract generally results in the lowest teleoperator utility due to its overly conservative strategy. Although DRL\_Contract achieves the highest teleoperator utility when the distributional shift is 0, it only surpasses DRO\_Contract by 2.25\%. However, as the magnitude of the distribution shift increases, DRL\_Contract's performance degrades significantly, reaching a 10.74\% lower teleoperator utility than DRO\_Contract in the worst case. Furthermore, as shown in Fig. \ref{fig12}, the type-8 ASP utility achieved by DRL\_Contract is 60.02\% lower than that of DRO\_Contract. Additionally, although the RO\_Contract achieves the highest ASP utility, the reward that teleoperators need to pay for AIGC service is overly high as well, which might discourage teleoperators from utilizing the AIGC service. Therefore, we can conclude that the contract bundle formulated via DRO\_Contract and SP\_Contract is robust in uncertain and varying AIGC service quality when the training data quality is satisfactory.

To evaluate the robustness of the proposed method against poor training data quality, we conducted additional experiments considering extreme points in the training data, where the teleoperator perceives the AIGC service quality as extremely poor. In Figs. \ref{fig9}–\ref{fig14}, we examine cases with 0, 50, and 100 extreme points, each set to a value of 1. As shown in Figs. \ref{fig10} and \ref{fig11}, the results closely resemble Fig. \ref{fig9}, demonstrating that DRO\_Contract is more robust than DRL\_Contract in maintaining teleoperator utility under varying levels of data distribution shift. Notably, SP\_Contract is less robust than DRO\_Contract when training data quality deteriorates and gradually degrades to RO\_Contract as data quality worsens.

In summary, upon preceding discussion, we can conclude that our proposed method DRO\_Contract is robust no matter of varying evaluation data distribution or training data quality, which is more suitable for formulating contract bundles in the case that AIGC services are uncertain.

\subsection{Convergence of DRO-based Contract Theory} \label{sec:6.4}
In this section, we analyze the convergence of our proposed DRO\_Contract under varying training data quality and present the results in Figs. \ref{fig15}, \ref{fig16}, and \ref{fig17}. The experimental results show that our proposed BCD-driven contract theory converges well across all types of contract bundles. Notably, the convergence speed of our method is primarily influenced by the number of contract bundles, $I$. The underlying rationale is that the monotonicity constraint guides the algorithm to first stabilize the leading contract bundles before optimizing the remaining ones. Furthermore, the converged latency values of the contract bundles also satisfy the monotonicity constraint in (\ref{P2_constraint_2}), demonstrating the feasibility of the proposed algorithm.

\section{Conclusion} \label{sec:7}
To facilitate the utilization of AIGC services in teleoperation, we designed a user-centric incentive mechanism from the teleoperators' perspective, thereby attracting more teleoperators to pay for AIGC services. However, information asymmetry, i.e., teleoperators possess limited information regarding the differentiated resource capacities of ASPs, and the uncertainty of AIGC service quality pose challenges to incentive mechanism design. To address this issue, we proposed a DRO-based contract theory framework, where contract theory was used to derive a differentiated reward scheme under information asymmetry, and DRO was integrated to handle uncertainty in AIGC service quality. With our proposed method, ASPs can efficiently formulate robust and user-centric incentive mechanisms, thereby ensuring the utility of both ASPs and teleoperators under varying conditions.

Notably, to solve the intractable DRO-based contract theory problem, we reformulated it as a tractable bi-level optimization problem using Propositions \ref{proposition_1} and \ref{proposition_2}, and then applied the BCD algorithm to iteratively derive robust contract bundles. Additionally, our proposed DRO-based contract theory extends traditional contract theory by enabling contract bundle formulation under uncertainty, which is generalizable to other problems. Simulation experiments demonstrated that our proposed method remains robust under varying training data quality and significant shifts in the evaluation data distribution. Moreover, the proposed method improved teleoperator utility by 10.74\% in the worst case and increased ASP utility by 60.02\% compared to DRL-driven contract theory.
\bibliographystyle{IEEEtran}
\bibliography{zhan}
\end{document}